%Paper: 9201036
%From: phr74ms%technion@taunivm.tau.ac.il
%Date: Sun, 19 Jan 92 19:46:33 IST

\input harvmac.tex
\def\RN{{{\hskip 3 pt}\rm l}{\hskip -4.7pt}{\rm R}}
\def\CN{{\hskip 3 pt}{\rm l}{\hskip -4.8pt}{\rm C}}

\def\half{{1 \over 2}}
\def\salf{{\textstyle{1 \over 2}}}
\def\ie{$i.\,e.$\ }
\def\to{\rightarrow}
\def\nub #1, #2, #3. { {\rm Nucl. Phys. }{\bf B#1} {\rm (#3) #2}}
\def\nubf #1, #2, #3, #4. { {\rm Nucl. Phys. }{\bf B#1 [FS#2]} {\rm (#4) #3}}
\def\plb #1, #2, #3.{ {\rm Phys. Lett. }{\bf #1B} {\rm (#3) #2}}
\def\lmpa #1, #2, #3. { {\rm Lett. Mod. Phys. }{\bf A#1} {\rm (#3) #2}}
\def\prl #1, #2, #3. { {\rm Phys. Rev. Lett. }{\bf#1} {\rm (#3) #2}}
\def\prd #1, #2, #3. { {\rm Phys. Rev. }{\bf D#1} {\rm (#3) #2}}
\def\pro #1, #2, #3. { {\rm Phys. Rev. }{\bf #1} {\rm (#3) #2}}
\def\jel #1, #2, #3. {{\rm JEPT Lett. }{\bf #1}{\rm (#3) #2}}
\def\jep #1, #2, #3. { {\rm Sov. Phys. JEPT }{\bf #1} {\rm (#3) #2}}
\def\cmp #1, #2, #3. { {\rm Comm. Math. Phys. }{\bf #1} {\rm (#3) #2}}

% For preprints in harvmac.tex
%\baselineskip=22pt plus 2pt
%\def\listrefs{\footatend\vfill\supereject\immediate\closeout\rfile\writestoppt
%\baselineskip=16pt\centerline{{\bf References}}\bigskip{\frenchspacing%
%\parindent=20pt\parskip=9pt
%\escapechar=` \input refs.tmp\vfill\eject}\nonfrenchspacing}

%Doublespace:
%\baselineskip=24pt plus 2pt

\def\inbar{\,\vrule height1.5ex width.4pt depth0pt}
\def\IB{\relax{\rm I\kern-.18em B}}
\def\IC{\relax{\hbox{{$\inbar\kern-.3em{\rm C}$}}}}

\def\ID{\relax{\rm I\kern-.18em D}}
\def\IE{\relax{\rm I\kern-.18em E}}
\def\IF{\relax{\rm I\kern-.18em F}}
\def\IG{\relax\hbox{$\inbar\kern-.3em{\rm G}$}}
\def\IH{\relax{\rm I\kern-.18em H}}
\def\II{\relax{\rm I\kern-.18em I}}
\def\IK{\relax{\rm I\kern-.18em K}}
\def\IL{\relax{\rm I\kern-.18em L}}
\def\IM{\relax{\rm I\kern-.18em M}}
\def\IN{\relax{\rm I\kern-.18em N}}
\def\IO{\relax\hbox{$\inbar\kern-.3em{\rm O}$}}
\def\IP{\relax{\rm I\kern-.18em P}}
\def\IQ{\relax\hbox{$\inbar\kern-.3em{\rm Q}$}}
\def\IR{\relax{\rm I\kern-.18em R}}
%\font\cmss=cmss10 \font\cmsss=cmss10 at 10truept%!!! should be 7pt
%\def\IZ{\relax\ifmmode\mathchoice
%{\hbox{\cmss Z\kern-.4em Z}}{\hbox{\cmss Z\kern-.4em Z}}
%{\lower.9pt\hbox{\cmsss Z\kern-.36em Z}}
%{\lower1.2pt\hbox{\cmsss Z\kern-.36em Z}}\else{\cmss Z\kern-.4em Z}\fi}
\def\IZ{\relax\ifmmode\mathchoice
{\hbox{Z\kern-.4em Z}}{\hbox{Z\kern-.4em Z}}
{\lower.9pt\hbox{Z\kern-.4em Z}}
{\lower1.2pt\hbox{Z\kern-.4em Z}}\else{Z\kern-.4em Z}\fi}
\def\IGa{\relax\hbox{${\rm I}\kern-.18em\Gamma$}}
\def\IPi{\relax\hbox{${\rm I}\kern-.18em\Pi$}}
\def\ITh{\relax\hbox{$\inbar\kern-.3em\Theta$}}
\def\IOm{\relax\hbox{$\inbar\kern-3.00pt\Omega$}}

\def \RN{\IR}
\def\half{{1 \over 2}}
\def\salf{\hbox{$\textstyle{1 \over 2}$}}
\def\balf{{{\bf 1} \over {\bf 2}}}
\def\ie{$i.\,e.$\ }
\def\CB{{\cal B}}
\def\CC{{\cal C}}
\def\CJ{{\cal J}}
\def\CL{{\cal L}}

\def \CN{{\cal N}}

\def\ub{\bar{u}}
\def\rh{\rho_{z}}
\def\ha{H_{z}}
\def\rb{\bar{\rho}_{\bar{z}}}
\def\hb{\bar{H}_{\bar{z}}}
\def\ch{\chi}
\def\chb{\bar{\chi}}
\def\pz{\partial_{z}}
\def\pzb{\partial_{\bar{z}}}
\def\FD{{\cal D}}
\def\pcz{\FD_{z}}
\def\pczb{\FD_{\bar{z}}}
\def\az{A_{z}}
\def\czb{\bar{C}_{\bar{z}}}
\def\cz{C_{z}}
\def\azb{\bar{A}_{\bar{z}}}
\def\JB{J^{\scriptscriptstyle B}}
\def\WS{\Sigma}
\def\rbp#1{\bar{\rho}_{\bar{z}}^{#1}}
\def\rap#1{\rho_{z}^{#1}}

\def\gin{g^{-1}}
\def\hin{h^{-1}}
\def\TR{{ \rm tr}}
\def\Appendix#1#2{\global\meqno=1\global\subsecno=0\xdef\secsym{\hbox{#1.}}
\bigbreak\bigskip\noindent{\bf Appendix #1 #2}\message{(#1 #2)}
\writetoca{Appendix {#1} {#2}}\par\nobreak\medskip\nobreak}
\def\re{\Re}
\def\im{\Im}
\def\vec#1{\overrightarrow {#1 \strut}}
\def\pr#1{\partial_{#1}}
\def\ket#1{\vert {#1}>}

\def\ghgh {\hbox{\raise.5ex\hbox{$ (G/H)$}$/$\lower.5ex\hbox{$ (G/H)$} }}
\def\nh{n}
\def\ng{N_g}
\def\n0{N}%sect. 5 level
\def\nk{K}%sect. 3 toratio

\Title{\vbox {\halign {# \hfil & \hfil #\cr
Technion & PH-34-90 \cr
TAUP & 1934 \cr}}}
{$G/G$ - Topological Field Theories by Cosetting $G_k$.}
\medskip
\centerline{{\bf Mordechai Spiegelglas }\footnote*{\hbox{
Lady Davis Fellow at the Technion. \hss {\tenpoint Bitnet:
phr74ms@technion.}}}}
%\smallskip
\centerline {Physics Department}
\centerline {Technion, Haifa 32000, Israel.}
\smallskip
\centerline {and}
\smallskip
\centerline{{\bf Shimon Yankielowicz}\footnote{**}{\hbox{
\hss{\tenpoint Bitnet:
h75@taunivm.}}}}
\centerline {Department of Physics and Astronomy}
\centerline {Tel-Aviv University, Tel-Aviv 69978, Israel.}
\noindent
\vskip 2cm

  {\abstractfont  $G/G$ topological field theories  based on $G_k$ WZW
models are constructed and studied. These coset models are formulated
as Complex BRST cohomology in $G^c_k$, the complexified level $k$
current algebra. The finite physical spectrum corresponds to the
conformal blocks of $G_k$ .The amplitudes for $G/G$ theories are
argued to be given in terms of the $G_k$ fusion rules. The $G_k/G_k$
character is the Kac-Weyl numerator of  $G_k$ and is interpreted as an
index. The Complex BRST cohomology is found to contain  states of
arbitrary ghost number. Intriguing similarities of $G/G$ to $c\leq 1$
matter systems coupled to two dimensional gravity are pointed out.}

\Date{}
%\draft
%For draft to work on with Shimon!!
%\baselineskip=22pt plus 2pt minus 1pt
%To save some timbers!!
%\baselineskip=13pt plus 2pt minus 1pt
%To save a timber!!
%\baselineskip=16pt plus 2pt minus 1pt
%\def\me{\hbox{\bf MS}}
\def\mei{{}}
\def\me{{}}

\newsec {Introduction and Summary.}

  Recently ``matrix models" were applied to the solution of two
dimensional gravity performing the sum over triangulations. This
converges to the sum over surfaces in the ``double scaling limit"
\ref\bk{E. Br\`ezin and V.A. Kazakov, {Phys.Lett.} {\bf 236B} (1990)
144.}$\, $\ref\ds {M. Douglas and S. Shenker, {Nucl.Phys.} {\bf B335}
(1990) 635.}$\, $\ref\gm {D. Gross and A.A. Migdal, {Phys.Rev.Lett.}
{\bf 64} (1990) 127.}. The amplitudes were found to obey the KdV flows
\ref\doug{M. Douglas, {Phys.Lett.} {\bf 238B} (1990) 176.}. The
amplitudes of topological two dimensional gravity \ref\witp{E. Witten,
Nucl.Phys. {\bf B340} (1990) 281.}$\, $\ref\dijw{R. Dijkgraaf and E.
Witten,  Nucl.Phys. {\bf B341} (1990) 167.}$\, $\ref\witr{E. Witten,
{\it Two Dimensional Gravity and Intersection Theory on Moduli Space},
IAS Preprint IASSNS-HEP-90/45 (1990).} were shown to obey similar
relations. The equality between amplitudes in matrix models  and in
topological gravity was explicitly verified as well, for small genera
\ref\jrnw{J. Horne Mod.Phys.Lett. {\bf A5} (1990) 2127.}. Thus, the
matrix models, while combinatorically performing the sum over all
surfaces, expose the topological features revealed by topological
gravity. In the framework of multi-\-matrix models one can describe
gravity coupled to $(p,q)$ minimal model, by going to higher critical
points in the ``double scaling limit" \ref\bdks{E. Br\`ezin, M.
Douglas, V.A. Kazakov and S. Shenker, {Phys.Lett.} {\bf 237B} (1990)
43.}$\, $\ref\gmi{D. Gross and A.A. Migdal, {Phys.Rev.Lett.} {\bf 64}
(1990) 717.}$\, $\ref\cgm{C. Crnkovi\'c, P. Ginsparg and G. Moore,
{Phys.Lett.} {\bf 237B} (1990) 196.}$\, $\doug. There are other ways
to discuss the coupling of matter to 2d gravity, like the traditional
Liouville approach \ref\nvil{N. Seiberg, Prog.Theor.Phys.Suppl. {\bf
102} (1990)  319 and references therein.}, in which the amplitudes
were recently calculated \ref\bek{M. Bershadsky and I.R. Klebanov,
Nucl.Phys. {\bf B360} (1991) 559.}$\, $\ref\igr{ I.R. Klebanov,  {\it
String Theory in Two-Dimensions} Trieste Lectures Spring 1991,
PUPT-1271 (1991).}, or the light-cone gauge approach implying
$SL(2,\IR)$ structure \ref\kpz{V. Knizhnik, A.M. Polyakov and A.B.
Zamolodchikov, {Mod.Phys.Lett.} {\bf A3} (1988) 819.}.

  Topological matter can be naturally coupled to topological 2d
gravity. An important example is given by the Topological Sigma Models
\ref\wits{E. Witten, Comm.Math.Phys. {\bf 118} (1988) 411.}. In this
work, we construct a new class of topological matter theories
associated with Rational Conformal Field Theories (RCFTs). We start
with theories based on current algebras, as those are believed to be
the basic building blocks of RCFTs \ref\ms{G. Moore and N. Seiberg,
Lectures on RCFT RU-89-32 (1989) and references therein.}. It should
be noted that  topological matter systems, which are not based on
CFTs, can be constructed and coupled to two dimensional (topological)
gravity, like the $CP^1$ sigma model in \witp. Here we focus,
nonetheless, on the derivation of topological field theories from
CFTs, a choice motivated by the possible insight into string theory
they can provide and even more, by their inherent simplicity. The
coupling  of topological CFTs to gravity becomes simpler due to the
vanishing trace of the energy momentum tensor, $T_{+-}=0$ ($T_{
\mu\nu}$ is just BRST exact in  the case of a general topological
field  theory \ref\dvv{R. Dijkgraaf, H. Verlinde and E. Verlinde,
Nucl.Phys. {\bf B352} (1991) 59.}).

  Rational Conformal Field Theories have been under intensive
investigation recently. Here we construct and analyze their
topological counterparts employing tools inherited from the study of
RCFTs \ms. While studying these topological theories we will unravel
the features of their ancestor RCFTs  which they encode, in particular
conformal blocks and their fusion rules. These features were
recognized as being among the essentials of RCFTs, starting with the
observation of E. Verlinde \ref\verl{E. Verlinde , Nucl.Phys. {\bf
B300} (1989) 360.} through the RCFT endeavor summarized in \ms. They
are also central to the profound relation which exists between RCFTs
(which are based on current algebras) and  the topological 2+1
dimensional Chern-Simons Gauge Theory \ref\csw{E. Witten,
Comm.Math.Phys. {\bf 121} (1989) 351.}.

  The coupling of these simple topological systems to (topological)
gravity, may give us a deeper insight into the structure of gravity
and into the way it couples to matter. {\sl What does gravity couple
to}, is an important question in the study of quantum gravity. $G/G$
topological theories provide interesting matter systems which enables
us to address this question  applying ideas and techniques developed
in the study of (R)CFTs. \me $G/G$ theories allow for the algebraic
study of two dimensional gravity.\foot{The Virasoro algebra, which
undoubtedly has an important r\^ole in the coupling of matter to 2d
gravity, is present in the universal enveloping algebra of the current
algebra and therefore, does not require any special attention.}
Recently, it was argued by Witten \ref\blkd{E. Witten, Phys.Rev. {\bf
D44} (1991) 314}\ (through a stringy black hole coset construction)
and emphasized by Eguchi \ref\egu{T. Eguchi, {\it Topological Field
Theories and the Space-time  Singularity}, Chicago, EFI and Tokyo
preprint, EFI-91-58, October 1991. }\ that space-time singularities in
general relativity have a natural description in terms of topological
field theories.

  In the present paper we construct the $G_k/G_k$ topological models.
The structure of these theories is intimately connected to the level
$k$ current algebra $G_k$ \ref\kac{V.G. Kac, {\it Infinite Dimensional
Lie Algebras} (Birkhauser, Boston, 1983).}$\, $\ref\gepw{D. Gepner and
E. Witten, {Nucl.Phys.} {\bf B278 (1986) 493.}} (based on the compact
Lie group $G$). In particular the physical spectrum of the $G_k/G_k$
topological model consists of the $G_k$ highest weight states. The
structure of $G_k/G_k$ is in fact inherited from the  corresponding
$G_k$ Wess$\, $\--Zumino$\, $\--Witten model \ref\wzw {E. Witten,
Comm.Math.Phys. {\bf 99} (1984) 455.} and the $G_k/G_k$ physical
states are in one to one correspondence with the conformal blocks of
the WZW model. Moreover, we provide evidence for our conjecture
\ref\giii{M. Spiegelglas and S. Yankielowicz, {\it Fusion Rules as
Amplitudes in $G/G$ Theories,} Technion PH-35-90 (1990)}\ that the
amplitudes in the $G_k/G_k$ topological theory are given by the fusion
rules of the corresponding $G_k$ WZW model. This was recently proven
by Witten \ref\witf{E. Witten, {\it  On Holomorphic Factorization of
WZW and Coset Models.} Princeton, Inst. Advanced Study preprint,
IASSNS-HEP-91-24 (1991).}.

  We start our investigation with the $G_k$ WZW model and construct
the $G_k/G_k$ topological model out of it. We follow the usual {\sl
gauged {\rm WZW} approach} to $G/H$ based on gauging the $H\subset G$
subgroup in the $G_k$ WZW model \me \ref\gko{P. Goddard, A. Kent and
D. Olive, Phys.Lett. {\bf B152} (1985) 88.}$\, $\ref\gkoc{P. Goddard,
A. Kent and D. Olive, Comm.Math.Phys. {\bf 113} (1987) 1.}$\,
$\ref\elci {K. Bardacki, E. Rabinovici and B. S{\"{a}}ring, Nucl.Phys.
{\bf B299} (1988) 151.}$\, $\ref\gak{K. Gaw{\c{e}}dzki and A.
Kupiainen, {Nucl.Phys.} {\bf B320} (1990) 625.}. By employing the BRST
variant of this construction \ref\karb{ D. Karabali and  H.J.
Schnitzer, {Nucl.Phys.}{\bf B329} (1990) 649.}$\, $\ref\kara{D.
Karabali {\it Gauged WZW Models and the Coset Construction of
Conformal Field Theories,} talk at NATO ARW 18th Int. Conf. on
Differential Geometric Methods in Theoretical Physics, Tahoe City, CA,
1989, Brandeis BRX-TH-275 (1989).} we  develope a BRST approach to the
$G/G$ models similar to the one Witten advocated for topological field
theories \ref\witdn{E. Witten, Comm. Math.Phys. {\bf 117} (1988) 353.}
and for topological  $\sigma$ models in particular \wits.  \me We
would like to impose the naive $G_k$ constraints, which turn out
however, to be second class constraints. This difficulty is overcame
by enlarging the theory, promoting it to a complex theory by adding
$G^c/G$ degrees of freedom. \me This extra $G^C/G$  degrees of freedom
form a WZW-like model with a coefficient $\bar{k}=k+ 2 c_G$, which can
be viewed as its ``level" ($c_G$ is the quadratic Casimir in the
adjoint representation of $G$). We denote it by $(G^c/G)_{\bar{k}}$,
noticing the formal analogy to the $G$ WZW model at level $-(k+ 2
c_G)$. We are, therefore, led to the study of the Complex BRST
cohomology of $G_k$. This constitutes {\sl the {\rm CBRST} approach}
to $G_k/G_k$.

  Complexification is also required in the {\sl the gauged {\rm WZW}
approach}. It turns out that in this approach, complex gauge
transformations are both natural and useful. Their gauge fixing
introduces a complex ghost system $(\rh,\ch)$ of spin  (1,0). Complex
gauge transformation allows us to set (locally) the field strength to
zero. Hence, they leave the stage clear to the moduli of the flat
gauge connections which are associated with the holonomies around
handles and holes (\ie\ insertions) in the {\sl world-sheet} $\Sigma$
\ref\atbt{M.F. Atiyah and R. Bott,  {Phil.Trans.$ $R.Soc.London}  {\bf
A308} (1982) 523.}. In both approaches $G^c$, the complexified version
of the group $G$, provides the natural setting for our investigation.

  The paper is organized as follows: In section 2 we construct the
$\IR/\IR$ topological model. Although it is based on a free scalar-\-$
$field in $\IR$ which is not a RCFT, it does have the virtue of
simplicity. We elaborate there on both the gauged WZW and the CBRST
approaches and demonstrate their equivalence. The abelian $\IR/\IR$
model contains the original scalar field $X$, an additional scalar
field $Y$ combining with $X$ into a complex scalar field and a complex
ghost pair $(\rh,\ch)$. This field content allows us to realize
$\IR/\IR$ as (the linear version of) the topological $\sigma$ model
\wits\ on $\IC$. \me Alternatively, $\IR/ \IR$ follows from the
additively gauged Lagrangian of the free scalar field $X$, upon
complexification and gauge fixing. \me The translational gauge
symmetry of $X$ can be fixed in this complex setting, eliminating the
propagating degrees of freedom of the theory. Fixing the $\IC$ gauge
algebra introduces the complex $(\rh,\ch)$ ghost system.

  Section 3 is devoted to the construction of topological theories
based on RCFTs. We start with a discussion of the spectrum, and
demonstrate that the result of the CBRST cohomology is equivalent to
the simple argument of \ref\witc{E. Witten, {\it The Central Charge in
Three\--$ $Dimen\-sions}, in {\bf Physics and Mathematics of String,
V. Knizhnik Memorial}, World Scientific (1990), p. 530.}\ describing
the $G/H$ fields as the fields of the $G$ theory which are primary
with respect to the chiral algebra of $H$. Both arguments prescribe
the $G_k$ highest weight states as the $G_k/G_k$ physical states
demonstrating that \me $G_k/G_k$ {\sl is the theory of $G_k$ conformal
blocks}. We then move to a simple RCFT of the rational torus or circle
of radius $r_{\nk}= \sqrt{\nk}$ and construct $U(1)_{\nk}
/U(1)_{\nk}$. In this $U(1)/U(1)$ case, CBRST is the more tricky
approach and we therefore, resort to the gauged WZW approach.
Associating physical operators with gauge holonomies around their
insertion points indicates that the amplitudes of $U(1)_{\nk}
/U(1)_{\nk}$ are given in terms of the $U(1)_{\nk}$ fusion rules. This
result is further investigated and generalized in \giii. We conclude
this section with the construction of $G_k/G_k$ for a general non-\-$
$abelian current algebra $G_k$. \me Gauging the WZW model according to
the Gaw{\c{e}}dzki Kupiainen approach \gak, to $G/H$ construction,
leads us to consider the complexified algebra $G^c$ in our $H=G$ case.
This recasts the action as a sum of three independent parts: a WZW
model for $G$ at level $k$ (analogous to $X$ in the $\IR$ case), the
$(G^c/G)_ {\bar{k}}$ WZW model with $\bar{k} =k+2 c_G$ (analogous to
$Y$ in the $\IR$ case) and the complex $(\rh, \ch)$ spin $(1,0)$ ghost
system in the adjoint representation. The $G^c$ WZW model produced,
has thus, two sets of bosonic currents. When properly supplemented by
the ghost contribution, they deserve the name ``complexified current
algebra''. Following the standard BRST approach we construct the BRST
operator $Q$. The currents $\CJ_z^a= [Q,\rh^a]_+$ satisfy the
Kac-Moody algebra with vanishing anomaly, $k^{{tot}}=0$. The energy
momentum $T_{zz}$ is also a BRST anti-\-commutator leading to
$c^{{tot}}=0$.

  In section 4 we investigate the mechanism of BRST cohomology applied
to current algebras. After discussing the cancellation of zero and
negative norm states in this complex BRST cohomology \ref\kaog{M. Kato
and K. Ogawa, {Nucl.Phys. } {\bf B212} (1983) 443.}$\, $\ref\huch{M.
Spiegelglas, Nucl.Phys. {\bf B283} (1987) 205 and talk at the XV Int.
Colloq. on Group Theoretical Methods in Physics, Philadelphia, 1986,
(World Scientific) p. 639.}, we turn to the calculation of characters,
employing the technique of ref. \gak. The $\lambda$ representation of
$G$ leads to a $G_k$ conformal block (all the representations we
discuss are integrable) whose Kac-Weyl \kac\ character is $\chi_
\lambda$. $G_k$ is then complemented by $(G^c/G )_{\bar{k}}$ along
with the ($\rho$, $\chi$) ghost system to generate $G_k/G_k$. We find
the $G_k/G_k$ character $M_{k,\lambda}$, to be given by the numerator
of $\chi_{k, \lambda}$. $M_{k,\lambda}$ plays the r\^ole of
``boundary" \ref\bndr{M. Spiegelglas, Phys.Lett. {\bf 247B} (1990)
36.} for the $\lambda$ representation\foot{$ M_{k,\lambda}$ contains
no states produced by current excitations. It is analogous to the Weyl
numerator serving as ``boundary" for a representation of a Lie
Algebra.} of $G_k$. $M_{k,\lambda}(q,u)$ ($q=e ^{2 \pi i \tau}$,
$\tau$ is the modular parameter for the torus and $u$, the moduli of
flat gauge connections on it) is useful in extracting two important
results about $G_k/G_k$. The torus partition function turns out to be
simply the number of $G_k$ conformal blocks ($k+1$ for the $SU(2)_k$
case). More detailed information follows by expanding $M_{k,\lambda}
(q,u)$ in powers of $q$. We interpret $M_{k,\lambda} (q,u)$ as ${\rm
Tr}\,(-1)^{\ng} q^{\n0}  e^{i \theta J_3}$ \ie\ the index or Euler
number for the BRST complex \huch. $\ng$ denotes the ghost number and
$N$ is the excitation level for the states  contributing to this
index, namely, states in the $Q$ cohomology. We will be able to read
the whole cohomology from $M_{k,\lambda}(q,u)$. We find \me infinitely
many states at all values of $\ng$, in addition to the $\ng=0$
physical state. Using an educated guess in order to construct the
cohomology, supplemented by demonstrating a correspondence between the
cohomology and algebraic relations in the $\lambda$ multiplet, we
argue that, in fact, the cohomology contains {\bf one} state at {\bf
each} $\ng$ sector. Those $\ng \not=0$ states are null states of the
$\lambda$ representation of $G_k$, ``dressed" by the ghosts and
$(G^c/G)_{\bar{k}}$.

  The BRST cohomology which we find, reveals some analogies to the
discrete states in two dimensional gravity, coupled to $c\leq 1$
matter. In the $c=1$ system, states in the Virasoro cohomology with
``non-\- physical'' ghost number, has been recently emphasized
\ref\disc{B. Lian and G. Zuckerman, {Phys.Lett.} {\bf 254B} (1991)
417; Yale preprint YCTP-\-P18-91 and references therein.}$\,
$\ref\wids{E.Witten, {\it The Ground Ring} Princeton, Inst. Advanced
Study preprint, IASSNS-HEP-91-24 (1991).}. \me There is also a
structural analogy between the $G/G$ system and two dimensional
Liouville gravity  The $G_k$ WZW model plays the r\^ole of the matter
system, while $(G^c/G)_{\bar{k}}$ plays the r\^ole of the Liouville
field, as hinted by general properties the two have in common.
$(G^c/G)_{\bar{k}}$ is pretty universal,\foot{This is apparent from
the expression of the $(G^c/G)_{\bar{k}}$ contribution to the
character \gak, which as seen in section 4, is independent of $k$ and
$\lambda$ apart from normalization.} it adapts to the $G_k$ ``matter"
system (resulting in $k^{\rm tot}=0$) and does not introduce
topological features of its own. \me The ghost system has the spin of
the symmetry generators, $(1,0)$ in the $G_k$ current algebra case and
$(2,-1)$ in the Virasoro gravitational case \ref\zam{We are grateful
to A~.B.~Zamolodchikov for discussion in this subject.}. Various
facets of this analogy between $G/G$ and two dimensional gravity are
pointed out and discussed throughout the paper.

  We conclude in section 5 with further remarks on the  $G/G$
topological theories. In particular, we put them in the context of
Landau-Ginzburg theories and 2+1 dimensional Chern-Simons theories. We
also briefly comment on their coupling to topological gravity in two
dimensions.

  In the appendix we review complex gauge transformations which are
an important tool in discussing the topology of gauge configurations
\atbt.

\newsec {Complex BRST Cosetting - the Abelian Case.}

  The $G/G$ theories are naturally formulated as gauge theories,
particularly, gauged WZW models in the non-abelian case. An
alternative formulation follows from the BRST gauge fixed variant. In
this section we introduce these two formulations for the abelian
$\RN/\RN$ theory. We start by twisting the free $N=2$ supersymmetric
theory ($c=3$) and subsequently present it as a Complex BRST (CBRST)
gauge fixed theory. This complex field theory consists of a complex
free scalar field $u(z)$ ($\IC$ being the target space), along with a
complex ghost system ($\rh$, $\ch$), of spin $(1,0)$. In this linear
version of the two dimensional $\sigma$ model \wits\ we will realize
that the CBRST mechanism is the gauge fixing of the translational
symmetry of $X(z)= {\rm Re}\bigl( u(z)\bigr)$. Once this symmetry is
identified,  we can write down  the gauge theory Lagrangian for our
model, \eqn\lgg{\CL_G=(\pz X - \az)(\pzb X-\azb)} taking the gauge
field $\az$ to be complex. By fixing this complex gauge algebra we
derive the same theory, as formulated in the CBRST approach. We thus,
show explicitly the equivalence of the two formulations, in the simple
case of $\IR/\IR$.

  Starting with the CBRST approach, the  $\IR/\IR$ case involves the
complex plane $\IC$ bosonic as target space for this twisted $N=2$
supersymmetric theory. Thus, the field $X(z) \in \IR$ is promoted to a
complex scalar field  $u(z) = X(z) + i Y(z)$ (and its complex
conjugate $\ub(z) = X(z) - i Y(z)$). We also introduce a complex ghost
system $\rh$ and $\ch$ of spins 1 and 0 respectively. $\rb$ and $\chb$
are their complex conjugates. The Lagrangian is \eqn\labl{\CL~
=~\salf\pzb\ub\,\pz u - i \rh\,\pzb\chb - i \rb\,\pz\ch} (the complex
structure of $u$ is thus, the symplectic form). The equations of
motion render $\rh$ and $\chb$ holomorphic whereas $\rb$ and $\ch$
turn out anti-\-holomorphic. The operator product expansion (OPE) are
\eqn\ope{\eqalign{ <\!\ub(z,  \bar{z})\,  u(w, \bar{w})\!>=\ln \vert
z-w \vert^2,\qquad & <\!\rh(z) \, \chb(w)\!>= {i \over (z-w)}\, , \cr
<\!\rb(\bar{z}) \, \ch(\bar{w})\!>&={-i \over (\bar{z}-\bar{w})}\,
.\cr }}

  In order to have a BRST symmetry, satisfying $\delta^2=0$  off-\-$
$shell, a complex spin 1 auxiliary field $\ha$ was introduced in ref.
\wits. The Lagrangian \eqn\lh{\CL_H~=~-\,2\,\hb\,\ha - \hb\,\pz u -
\ha\,\pzb \ub - i \rh\,\pzb\chb - i \rb\,\pz\ch} is invariant under
the BRST transformations $\delta u = i \ch$, $\delta \ub = i \chb$,
$\delta \rh = \ha$, $\delta \rb = \hb$ and $\delta \ch = \delta \chb =
\delta H = 0$. $\CL_H$ is in fact a BRST anti-\-commutator \eqn\topl{
\CL_H~= ~\delta\, (-\, \rb\, \pz u - \rb\,\ha - \rh \,\pzb \ub -
\rh\,\hb)} and is hence a topological Lagrangian. We will work
on-shell, since holomorphic fields on the world sheet are sufficient
when we consider the current algebra and the energy momentum tensor.
Then, $\delta \rh = \pz u$ and  $\delta^2=0$ follows with the use of
the $\ch$ equation of motion. The energy momentum tensor is also a
BRST anti-\-commutator \eqn\tenr{ T_{zz}~=~\pz u \, \pz \ub + i \rh \,
\pz \chb~=~\delta b_{zz}.} $b_{zz}=\rh\,\pz \ub$ is the fermionic
partner of $T_{zz}$ and also the Virasoro ghost as the fermionic
partner of $T_{zz}$.

  The BRST transformation $\delta$ is generated by the BRST operator
$Q= \oint\! \JB_{z} \!dz$ with the BRST current given by $\JB_{z}=\chb
\, \pz u$.  Calculating the cohomology of $Q$ we impose the
constraint $\CJ_{z}(z) = \pz u = 0 $, which is a first class
constraint since it is a BRST commutator. \me This makes the reason
for complexification now apparent. Starting with a real free scalar
field Lagrangian, $\CL_X= \pz X \pzb X$, we could naively try to
impose the current $j_z(z)= \pz X$ as the constraint. However, it is
not a first class constraint (since $<\!j_z(z) \, j_z(w)\!>=
{1/(z-w)^2}$). We resort to complexifying $X$ and impose the current
$\CJ_z$ encouraged by $<\!\CJ_z(z) \, \CJ_z(w)\!>= regular$. By adding
ghosts, we get a twisted $N=2$, supersymmetry or a BRST symmetry and
can employ BRST cohomology to impose $\CJ_z=0$. This algebraic
argument shows that the topological theory in hand, is actually a
scalar field theory, which has its translational symmetry $X \to X +
\alpha$ gauged and fixed by complexification. In short, these symmetry
considerations imply that this theory, presented so far as a $\sigma$
model, is actually the gauged $\RN$ model which is identified with
${\RN}/{\RN}\,$.

  Let us show explicitly the equivalence of the CBRST {\sl approach}
presented so far, to the more conventional Gauged Wess-Zumino-Witten
model {\sl approach} for coset construction \gko$\, $\gkoc$\,
$\elci$\, $\gak. Additively gauging $\CL_X$ gives: \eqn\lg{
\CL_G~=~(\pz X - \az)\;(\pzb X -\azb).} We are now interested in
complex gauge transformations, $\delta \az= \pz \lambda$, $\delta
\azb= \pzb \bar{\lambda}$ and $\delta X = i {\rm Re}\,\lambda$.
Although the action \lg\ is only invariant under real gauge
transformations, there are two reasons for the use of complex gauge
transformations. First, complex gauge transformations can be used to
set the field strength $F= \pz \azb + \pzb \az $ to zero (for a
compact world-sheet $\WS$, a necessary condition would be the
vanishing of the first Chern class $c(F)= \int_{\WS}\! F\, ds ~$
\giii$\, $). This keeps the stage clear for the topology of the gauge
configurations, \ie holonomies around handles and holes of $\WS$.
Hence, complex gauge algebra provides a tool for studying topological
aspects of gauge configurations \atbt. This is discussed further in
the appendix. The second reason to employ complex gauge
transformations is that they are required to fix the gauge properly,
by CBRST techniques. Unable to impose $\pz X = 0$, we resort to
complexifying the currents along with the gauge fields coupled to
them.

  For a topologically trivial $\WS$ ($\IC$ or a disk), we can see two
ways to trade $\az$ for the scalar field $Y$. First, considering only
real gauge transformations (which are the actual symmetries of the
action) we make the change of variables $\az=i \pz Y$, $\azb= -i \pzb
Y$. The $\rho$, $\chi$ ghost system accounts for the Jacobian
resulting from the change in the functional integration measure \gak.
Alternatively, we consider complex gauge transformations with $\az=i
\pz Y$, $\azb= -i \pzb Y$ as the gauge condition. Since the action is
not invariant under all the complex gauge transformations it does
depend upon the gauge choice $Y$. In particular, the action depends on
the complex transformations that change the curvature and are exactly
described by the field $Y$. In other words, $Y$ accounts for the
curvature, the local information encoded in the gauge configuration
(see the appendix for more details). From this point of view $\rh$ and
$\ch$ are the Fadeev-Popov ghosts associated with this complex gauge
fixing. Either ways \labl\ follows from \lg.

  We have formulated the $\RN/\RN$ coset model both as  a gauged
abelian WZW model and as a topological theory in the CBRST approach.
In the latter, complex formulation for both matter and ghost systems
is required. The need for a complex formulation is less apparent when
$\RN/\RN$ is written as a gauged WZW model. However, in order to avoid
a gauge anomaly, complex gauge algebra is the proper way to BRST
quantize the model. This promotes the bosonic degrees of freedom to a
complex scalar. Gauge fixing requires a complex ghost system showing
that the two formulations are indeed equivalent.

  The next section leads us to the nonabelian case and following \gak\
we will define the $G/G$ coset model as a gauged WZW model. The gauge
symmetry is fixed yielding a non-\-abelian generalization of \labl.
The result involves $G^c_k$ and may be formally described as a new
kind of a  $G^c_k$ WZW model with complex ghosts and current algebra.
In the  nonabelian case $G^c_k$ takes the r\^ole of $\IC$ in the
$\RN/\RN$ case.

\newsec { BRST Cosetting in Rational Field Theories.}

  In the previous section the abelian coset model $\RN/\RN$ was
recasted as a topological 2d theory, formulated in complex BRST terms.
We would now like to discuss non-abelian theories as well. We thus
proceed to discuss Rational Conformal Field Theories having finite
number of highest weight states. We concentrate on the cosetting of
the level $k$ WZW model which has the Kac-Moody algebra $G_k$ as its
chiral algebra. The highest weight states of $G_k$ are of particular
interest. As we will soon argue, the physical spectrum of the $G/G$
theory consists exactly of these $G_k$ highest weight states \witc. We
will then turn to a detailed construction of the simplest rational
$G_k/G_k$ model, namely the topological model of $U(1)_K/U(1)_K$,
following from the abelian RCFT of the rational torus. We will dwell
on some of its peculiairities and then spend the rest of this section
on the construction of the general non-abelian $G_k/G_k$ theory.

\subsec {The Spectrum of $G/G$.}

  We wish to determine the spectrum of the $G/G$ model by imposing the
$G_k$ constraints via BRST cohomology. It is useful to compare the
BRST approach to a simpler argument for the spectrum of coset models.
We are restricting the scope of the discussion to the holomorphic part
of the theory (We can always glue it to the anti-holmorphic part and
get the diagonal modular invariant. Other invariants are beyond the
scope of our discussion here, {\it c.f.} \ref\lgf{M. Spiegelglas, {\it
Setting Fusion Rules in Landau-Ginzburg Space} Technion pre\-print
PH-8-91, Phys.Lett.{\bf B},  to appear}). The operator content of the
coset model is then easily determined. $G/H$ contains the operators in
the $G$ theory, which are primary with respect to the chiral algebra
of $H$ \witc\ (in ref. \witc\ this  observation serves as a definition
of the $G/H$ model and could be easily generalized to $(G\!/\!H)\,
/\,(G\!/\!H)$ as will be discussed in section 5\thinspace).

  Let us show how this spectrum of $H$ highest weight states in the
$G$ theory follows in the language of BRST cohomology. The constraints
we should impose are $j_z^H$, the currents of $H$. As explained in the
$\IR/\IR$ case, these constraints are second class and can not be
imposed quantum mechanically. As in the abelian case, first class
constraints $\CJ_{z}^H(z)$ follow from $j_z^H$,  once extra bosonic
degrees of freedom are added. These $H^c/H$ degrees of freedom were
introduced by \gak\ and further studied by Karabali and Schnitzer
\karb. They give rise to currents which combine with $j^H_z$ into the
complex  currents $\CJ_{z}^H(z)$ and $\bar{\CJ}_{z}^H(z)$. $\CJ_{z}
^H(z)$  are the constraints imposed by the BRST cohomology. These
complex currents also contains a contribution from the spin $(1,0)$
($\rh$, $\ch$) ghost system, as expected in a non-abelian algebra of
constraints. Quantum mechanically, by imposing the current algebra
$\CJ_{z}^H$ via the  cohomology of the BRST operator  $Q_H= \oint dz
\,\ch \,\CJ_{z}^H + a~ghost~part\,$\foot{ $Q_H$ contains a ghost part,
although we have included ghosts in  $\CJ_{z}^H(z)$, since $Q_H$
requires only $\half \ch$ times the ghost part of the currents.}, we
cause the negatively indexed modes of the  current $\CJ_{z}^H$ (as
well as those of the complex conjugate $\bar{\CJ}_ {\bar{z}}^H$) to
vanish on the physical Hilbert space. As a matter of fact, the
physical Hilbert space is exactly the $\ng=0$ sector of this CBRST
cohomology and is free of all the $H$ excited states. It is only due
to the  zero modes of $H$. In section 4 we will explicitly see the
cancellation of the $H$ excitations in pairs and find the cohomology.
Therefore, the physical spectrum consists of the $H$ highest weight
states in the $G$ theory. The CBRST has removed $H$ descendants from
the cohomology producing the expected $G/H$ spectrum.

  The same procedure applies to the construction of $G/G$ theories.
Imposing the chiral algebra $G_k$ leaves only the highest weight
states of the $G_k$ theory in the physical Hilbert space. It should be
noted that all the highest weight states (or the zero modes which give
rise to them) are inherited from the original $G_k$ theory and are not
produced through the CBRST construction. The extra $H^c/H$ and ghost
degrees of freedom, added in the CBRST constructing of $G/H$ (and in
particular $G^C/G$ which is added in the $G/G$ construction), serve to
eliminate the $H$ descendant states.\foot{The (C)BRST mechanism
eliminates states in ``splitted-pairs" as we will discuss in section
4. In $G/H$ the pair is splitted between ${\CJ}_ {\bar{z}}^H$ and
$\bar{\CJ}_ {\bar{z}}^H$ excitations.  Therefore, on the cohomology,
the excitations of ${\CJ}_ {\bar{z}}^H$ and $\bar{\CJ}_ {\bar{z}}^H$
vanish as well as those of $j_z^H$  and the $H^c/H$ currents.} They do
not contribute extra states to the spectrum. Hence, the topological
structure exposed in the $G/G$  topological field theory resides in
the original $G_k$ theory and not in the topologically trivial $G^c/G$
extension. The $G/G$ theory is the theory of the $G_k$ conformal
blocks. It is therefore of little surprise that $Z$, the torus
partition function in this theory just counts those conformal blocks.
The amplitudes turn out to be generated from the $G_k$ fusion rules
\giii, in analogy with the $S^2 \times S^1$ Chern-Simons amplitudes
\csw.

\subsec {The Rational Torus.}

  Although the $\RN$ model of the previous section is not a RCFT, it
could still be claimed that the spectrum of the $\RN/\RN$ model
consists of the highest weight states \ie\  the Fock ground states,
given as eigenstates of the momentum $p$. The $U(1)_{\nk}$ case of the
rational torus provides us with an instructive debut into the
construction of topological theories based on RCFTs. Its formulation
relies on flat $U(1)$ gauge connections \gak\ and hence differs from
the CBRST formulation, which is applied in the subsequent nonabelian
case.

  The Lagrangian $\CL$ is given by \lg \eqn\lgp{\CL~=~(\pz X -
\az)\;(\pzb X -\azb).} \nref\mcm{G. Moore and N. Seiberg, Comm. Math.
Phys. {\bf 123} (1989) 177.}However, $X$ is now periodic and we choose
its period to be ${2 \pi \over \sqrt{\nk}}$. The possible $\az$
holonomies are then ${2 \pi n \over \sqrt{\nk}}$, exactly like flux
quantization in a superconductor or in the presence of a charge
$\sqrt{\nk}$ Higgs field. So far, everything is independent of the
period of $\az$ itself. Choosing $\az$ as a $U(1)$ field of $ {2 \pi
\sqrt{\nk}}$ period\foot{The normalization of the fundamental charge
is ${1 \over \sqrt{\nk}}$, chosen to comply with the standard
presentation of the Rational Torus \mcm\ with $\nk$ states.}, makes
its holonomies close into a $Z_{\nk}$ group, since those are now
additive modulo $\nk$. $\nk$ is an integer and we are actually dealing
to with the Rational Torus Model with radius $r_{\nk}= \sqrt{ \nk
}\,$. In other words, the possible holonomies around a contour $\CC$
are $ H[\az^n;\CC]= \oint_{\CC} \vec{A^n} \cdot d \vec{l} = {2 \pi n
\over \sqrt{\nk}}$, $0<n \leq \nk$ and $\az^n$ denotes a gauge
configuration with the holonomy ${2 \pi n \over \sqrt{\nk}}$. These
$\nk$ possible holonomies correspond to the $\nk$ primary fields in
the  $r_{\nk}$ rational torus model \mcm.

  Indeed, considering the construction of amplitudes in the
$U(1)_{\nk}/U(1)_{\nk}$ model \giii, we find the possible insertions
to consist of the ${\nk}$ holonomies presented above. Moreover, the
spectrum of the ${\nk}$ possible insertions or vertex operators in the
$U(1)_{\nk} /U(1)_{\nk}$ model, along with their $Z_{\nk}$ structure,
corresponds to the primary operators of the $r_{\nk}$ rational torus
in agreement with the discussion we had in subsection (3.1).

  Introducing operators by holonomies around the insertion points
raises an interesting issue. In order to calculate a Wilson loop
observable $W_q[\az;\CC]$, we should calculate the exponent of a
holonomy, $W_q[\az;\CC]= \exp ( iq H[\az;\CC])$. This is the ``trace"
in the $q$ representation of $U(1)$, with the charge $q/\sqrt{\nk}$
running around the Wilson loop $\CC$.  $W_q[\az^n;\CC]=  S_{nq}=e ^{{i
2\pi  \over {\nk}}{n q}}\,$ is the transformation matrix between two
bases in our Hilbert space of $U(1)_{\nk}$ conformal blocks. $S_{nq} $
transforms the basis spanned by $\ket {n}$ with $n$ specifying the
inserted flux at the puncture; into the basis spanned by $\ket{q}$
specifying the charge (which is used to probe the insertion in the
trace leading to $W_q[\az;\CC]$). The transformation matrix $S_{nq}$
is actually the Hartle-Hawking wave function generated by the
insertion of the operator $n$ into the path integral, in the $\ket{q}$
basis (specified on $\CC$). $S_{nq}$ is also the matrix representing
the modular transformation $S$ ($\tau \to -1/\tau$) on the world-sheet
torus, for the original $U(1)_{\nk}$ model. In \giii\ $S_{nq}$ is used
to calculate the amplitudes in this model (the arguments based on
\csw\ apply to general $G/G$ models as well). The amplitudes,
following from $S_{nq}$, are given by products of the $U(1)_{\nk}$
fusion rules.

  It should be noted that we have formulated the $U(1)_{\nk}/
U(1)_{\nk}$ case and determined the spectrum through studying the
topology of the $U(1)$ field configurations, while we avoided gauge
fixing. Now we are going to attempt to gauge fix $U(1)_{\nk}/
U(1)_{\nk}$, via CBRST and point out the difficulties. We try to use
the $\rho$, $\chi$ ghosts to fix the $U(1)$ gauge symmetry, expecting
$\CL={\salf\pzb \ub\,\pz u - i \rh\,\pzb\chb - i \rb\,\pz\ch}$, like
in \labl, to be the gauge fixed Lagrangian. A puzzling problem arises
concerning the periodicity appropriate for $Y$. The options seem to be
${2 \pi \over {\sqrt{\nk}}}\,-\,$the period of the field $X$ and ${2
\pi\sqrt{\nk}\,-\,}$the period of $\az$ gauge fixed by $Y$. However,
it turns out that no single field $Y$ is sufficient.

  In the CBRST cohomology the ghosts $\rho$, $\chi$ serve to eliminate
the excitations created by the $X$ and $Y$ oscillators. They leave the
spectrum, which is associated with the zero modes of $X$ \ie\  the
(quantized) momentum and winding number of $X$. This infinite set of
states is in contradiction with the expected finite dimensional
Hilbert space. The latter space contains $\nk$ states, in
correspondence with the $\nk$ primary states of $U(1)_{\nk}$.\foot{It
is well known \ms\ that the $U(1)_{\nk}$ primary states are associated
with the momentum values taken modulo the winding. This follows from
the exponential operators $M_{\pm}=e^{\pm iX \sqrt{{{\nk} }}}$, which
together with the momentum current $\pz X$, generate the  $U(1)_{\nk}$
chiral algebra. The possible primary operators are, therefore, $e^{i
p_n X}$ with $p_n= {n \over \sqrt {{ {\nk}}}}$. Only $-{{{\nk} \over
2}} <n \leq  {{{\nk} \over 2}}$ yield states which are also primary
under $M_{\pm}$.} We conclude, therefore, that the cancellation of
more states is required in order to get a finite dimensional BRST
cohomology. More ghost states are needed, which could pair with the
extra momentum states and cancel in quartets \kara$\, $\karb. However,
we seem to have used up all the symmetry of the $U(1)$ theory  with
all the ghosts it implies and still were not able to produce the
expected finite dimensional Hilbert space of $U(1)_{\nk}/U(1)_{\nk}$.

  Looking for a resolution for this puzzling situation of missing
symmetries and ghosts, we focus for a moment on the ${\nk}=2$ case.
The equivalence of $U(1)_2$ to $SU(2)_1$ gives us a clue. While
imposing constraints we have overlooked the $j^{\pm}_z$ part of the
chiral algebra $SU(2)_1$ $\bigl($and for ${\nk}\not=2$ we did not
impose the $M_{\pm} = \exp(\pm i {\sqrt{ {\nk}} X})$ operators in the
chiral algebra$\bigr)$. Extra ghosts, $\rho^{\pm}$ and $\chi^{\pm}$,
would be necessary to fix this additional part of the chiral algebra.
Moreover, the bosonic degrees of freedom were also not properly taken
into account. Although $SU(2)_1$ is easily bosonized and described by
a single periodic scalar, its complexification is not readily
described this way. Thus, we have to complexify $SU(2)_1$ without
relying on its realization as a scalar field on a circle. Three
additional scalar fields are, therefore, required to promote $SU(2)_1$
to $SL(2,\IC)$ (as discussed later in this section) rather than a
single scalar $Y$. The full set of the extra bosonic degrees of
freedom needed to complexify the current algebra oughts, therefore, to
be identified in order to proceed with a CBRST formulation of
$U(1)/U(1)$. The addition of a single field does not seem sufficient,
even though the original $SU(2)_1$ (or $U(1)_{\nk}$) can be formulated
in terms of a single periodic field. The general  $U(1)_{\nk}$ case is
harder to formulate using CBRST, since the chiral algebra contains
operators of spin ${\nk}$ that result in ghosts with varying spins. In
this $U(1)_{\nk}$ example, the gauged WZW approach seems simpler. We,
therefore, leave the proper complexification of the rational torus
along with the corresponding CBRST approach, for a future work.

  We would like to point out that our attempt above at applying a
CBRST construction for $U(1)_\nk$ gives rise to a RCFT which is
interesting in its own right and also in the context of two
dimensional gravity. We managed to cancel the oscillator excitations
in the rational torus. We were left with the excitations due to the
quantized values of the momentum. Those values were not bounded by the
winding convincing us to dismiss this (C)BRST construction for
$U(1)/U(1)$. By a slight abuse of notation, we could denote the theory
we have found, of $X$ primaries in $U(1)_{\nk}$, by $U(1)_{\nk}/\IR$.
It is  different from the topological theories we have discussed so
far, since it has an infinite dimensional (although discrete) Hilbert
space.

  For the $U(1)_2=SU(2)_1$ case, this CBRST construction yields the
discrete $c=1$ Virasoro states. This follows upon the examination of
the $X$ primary states in $SU(2)_1$. Then, the application of
$J_X^{\pm}= \oint j^{\pm}_z dz$ on these states, accounts for all the
Virasoro primaries in the $X(z)$ Fock space\foot{A short
demonstration: In the one dimensional realization of $SU(2)_1$ where
$j^3_z= \pz X$, the $X$ primaries are also Virasoro primaries. Upon
the application of $J_X^{\pm}$ those states stay Virasoro primaries
(since the Virasoro generators are scalar under the global $SU(2)$
algebra, generated by the zero modes of the $SU(2)_1$ currents
including $J_X^{\pm}$). We thus have produced Virasoro primaries in
the Fock space created by the oscillators of $X$.} or equivalently the
zeros of the Kac determinant (with the right degeneracy). Recently,
the discrete Virasoro states have drawn much attention \disc$\,
$\wids, especially those generating the ground ring for $c=1$ two
dimensional gravity. Here they are shown to constitute the $SU(2)/\IR$
modules. This observation merits further study. It reveals an
additional facet of the relation between the WZW model and two
dimensional gravity, a theme which is developed throughout the paper.

  The discussion above is easy to generalize to tori in higher
dimensions. If we choose a torus generated by the root lattice of the
simply laced simple Lie algebra $G$, we will have the interesting
description of the topological model $G_1/G_1$ using abelian gauge
fields. These cases could also be described in the CBRST formulation
following from the nonabelian description of $G_1/G_1$  which is the
subject of the coming subsection. \me It should be noted, that in this
topological context (or in the CS context), we have a theory with both
an abelian and a nonabelian description.

\subsec {The Nonabelian Case.}

  We would like now to construct the $G/G$ theory by imposing the
currents of a non-abelian Kac-Moody algebra $G_k$ as constraints on
the $G_k$ theory. Like in the abelian case, we will employ a complex
ghost system $\rh$, $\ch$ and $\rb$, $\chb$, taken in the adjoint
representation of $G$. From $\IR/\IR$ we know that we ought to
complexify the Kac-Moody currents, prior to imposing them, since the
$G_k$ currents are not first class for $k\not=0$. The extra bosonic
degrees of freedom required to ``complexify the model", are found
through the gauged WZW model. The WZW action is \wzw\ \eqn\swzw{
S_k(g)~ =~ {k \over 8 \pi} \int_{\WS} \!\TR\, (\gin \pz g \, \gin \pzb
g) ~+~ {k \over 12 \pi} \int_{\CB} \TR\, (\tilde{g}^{-1} d \tilde{g}
)^3,} $\tilde{g}$ extends $g$ into $\CB$, such that $\partial \CB
={\WS}$.

  In \gak\ the $G/H$ coset model was formulated as the gauged WZW
model \eqn\gwzw{S_k (g;\!A)\,=\, S_k(g)+{k \over 4 \pi} \TR\!
\int_{\WS}\!\! (\az \,\pzb g \gin \!-\azb \, \gin \pz g + \az\, g \azb
\gin \!- \az \, \azb),}  where $g \in G$ and the (anomaly free)
subgroup $H \subset G$ is gauged. We will be interested in the case
where $H=G$. For $\WS$ topologically trivial (a disk or $\IC$ with no
insertions), we can choose the gauge $\az = i\, h^{-1} \pz h$, $\azb =
-i\,h^{*} \pzb {h^{*}}^{-1} $. The complex gauge transformation $h(z)
\in G^c$ takes into account local features of the gauge configuration,
namely, curvature or holonomies around shrinkable loops. When the
world-sheet $\WS$ is compact there is the restriction of trivial first
Chern class, or else there is a global obstruction for solving the
equation $\az = i\,h^{-1} \pz h$.\foot{In the abelian case the
corresponding statement is that solving the Poisson equation for the
gauge condition, with the magnetic field $F$ being the source,
requires that the monopole number is zero.} We restrict ourselves to
gauge configurations with zero first Chern class and overlook this
restriction as long as we are working locally.

  Following \gak\ we change variables from $\az$ to $h$. Among the
complex gauge transformations $h(z) \in G^c$ the unitary
transformations are really gauge symmetries and $S_k(g;A)$ is
independent of them. $S_k(g;A)$ only depends on $hh^{*}$ and the
Polyakov-Wiegmann formula gives \ref\pw{A.M. Polyakov and P.B.
Wiegmann, {Phys.Lett.} {\bf 131B} (1983) 121.} \eqn\sbo{S_k(g;A) ~=
{}~\,S_k(g) ~-~S_k(hh^{*}),} which seems like the required
``complexification". In the second term, $h$ can be taken in
$G^{c}/G$. The bosonic action is therefore, a WZW action for the group
$G^{c}$. From the first term one derives the original $G_k$ currents
while the second term is producing their $G^c/G$ counterparts. \sbo\
thus provides the $G^c$ current algebra.

  Further evidence for complexification in $G/G$ theories is found for
$G=SU(N)$ through bosonization \ref\denm{We are grateful to D.
Nemeschansky for collaboration in this direction.}\ \me (which offers
an additional approach to the study of $G/G$). Expressing the currents
in terms of bosons,\foot{Some of the bosons can be replaced by the
appropriate $(\beta,\gamma)$ bosonic ghost system of spin $(1,0)$.} it
is straightforward to check that bosonic fields, originating from the
two terms, naturally combine into complex fields.

  The ghost system is introduced by the Fadeev-Popov determinant upon
gauge fixing or, alternatively, as the Jacobian resulting from the
change of variable from $\az$ to $h(z)$. The ghost field $\chi$ is a a
fermionic parameter for the infinitesimal gauge transformation. The
anti-\-ghost $\rho$ has the following origin (in analogy with an
argument for the Virasoro anti-\-ghost of the bosonic string
\ref\grsw{J. H. Schwarz, M. B. Green and E. Witten, {\it
Superstrings}, Cambridge Press (1987), Chapter 3.}). Topological
reasons may prevent having $\az = i\,h^{-1} \pz h$. We then settle for
the closest $\az$, with minimal difference from $i\,h^{-1} \pz h$.
Thus, we minimize the ``action'' \eqn\smin{{\cal S}~=~\TR\,(\az -
i\,h^{-1} \pz h)\,(\azb + i\,\pzb h^{*} {h^{*}}^{-1}).} The
anti-\-ghost $\rh$ denotes the difference $\az - i\,h^{-1} \pz h$
(more precisely, this difference is $\ha$, the bosonic counterpart of
$\rh$, introduced in section 2).  It satisfies the Euler-\-Lagrange
equation $\pczb \rh=0$, $\pczb = \pzb +\azb$. The zero modes of $\rh$
are associated with the moduli of the flat gauge configuration
corresponding to $\az$ (via a complex gauge transformation). \me This
argument, which actually determines the classical degrees of freedom
left in the theory (we will soon argue that non-\-zero modes are
canceled), has its counterpart for the bosonic string. The zero modes
of the  Virasoro anti-ghost $b$ in the bosonic string case are the
moduli of the complex structure on the Riemann surface. They are given
by the two dimensional metric up to diffeomorphisms. In our case, the
gauge moduli, or the zero modes of the anti-\-ghosts, correspond to
gauge configurations up to complex gauge transformations.

  With the ghosts included, the action for the $G/G$ model is
\eqn\stot {S_k(g;h,\rho,\chi) ~=~\, S_k(g) \,-\, S_k(hh^{*}) \,-\, i
\int_{\WS} \! \TR\, \Bigl( (\rh \,\pczb \chb ) \,+\,  (\rb \,\pcz \ch
)\Bigr).}  To get rid of the remaining $\az$ dependence in $\pcz$ and
$\pczb$, we perform a gauge transformation on the determinant which
gives rise to the chiral anomaly $\exp\bigl(S_{2c_G}(hh{*})\bigr)$.
$c_G$ is the Casimir of the adjoint representation of $G$ or its dual
Coxeter number. We thus get \eqn\stol{S_k(g;h,\rho,\chi) ~=~\, S_k(g)
\,-\, S_{k+2c_G}(hh^{*}) \, -\, i\! \int_{\WS} \!\! \TR\, \Bigl( (\rh
\,\pzb \chb ) + (\rb \,\pz \ch )\Bigr).} We see that actually $S_k(g)
\,-\, S_{k+2c_G} (hh^{*})$ should be considered as the ``complexified"
WZW action for $G^c$. The significance of the particular coefficient
in $S_{k+2c_G}(hh^{*})$ will be discussed shortly.

  Let us suggest a convenient gauge choice for $h \in G^c$. This
relies on the fact that  the action $S_k(g;  h,\rho, \chi)$ depends
only on the combination $hh^*$ and thus, $ h$ can be taken  in $
G^c/G$. We can multiply $h(z)$ by $f(z)$, a (unitary) element of $G$.
This is a gauge transformation that will not change the action. We can
choose $f(z)= \sqrt{h(z)h(z)^{*}} h(z)^{-1}$ with $f(z)h(z)$
hermitian. A hermitian field $h(z)$ is our gauge choice in analogy to
$\az = i\pz Y$, in the abelian case (where $Y$ is real). For
topologically non-trivial $\WS$, phases associated with the
non-trivial holonomies, should be introduce. We will sum over these
holonomies in the next section, where $\WS$ is the torus. We will
further discuss holonomies and the gauge choices associated with them
in \giii.

  Let us now demonstrate that we have really constructed a topological
matter system by presenting the energy momentum tensor $T_{zz}$ as a
BRST commutator, as we did in the abelian case. Again we find  the
on-shell version of the BRST transformations sufficient for the
discussion of the chiral part of the model. The currents are: $j_z = k
\gin \pz g$, $j_{\bar {z}} = - k \pz g \gin $ and $i_z = -(k+2c_G)
\hin \pz h$, $i_{\bar{z}} = (k+2c_G) \pzb h \hin $. We can now
construct the the bosonic parts of the complexified currents
$\CJ_{z}=j_z +i_z$, $\bar{\CJ}_{z}=j_z -i_z$, $\CJ_{\bar{z}}=
j_{\bar{z}} -i_{\bar{z}}$ and $\bar{\CJ}_{\bar{z}}= j_{\bar{z}}
+i_{\bar{z}}$. The contribution of the nonabelian ghosts to the
currents (written momentarily with indices, to display the structure
constants $f^{abc}$) is ${J_{{z}} ^a} ^{gh}= i f^{abc} \rap{b}
\chb^{c}$, ${\hbox{$\bar{J}_ {\bar{z}}^a$}} ^{gh} = i f^{abc} \rbp{b}
\chi^{c}$. The chirality of the ghost system leaves $\bar{\CJ}_{z}$
and $\CJ_{\bar{z}}$ with no ghosts contribution. This reveals an
interplay between ``world-sheet'' and ``space-time" holomorphicity,
which is typical to $N=2$ supersymmetric two dimensional theories
(playing an important r\^ole in (2,0) constructions \ref\huw{C. Hull
and E. Witten, Phys.Lett. {\bf 160B} (1985) 398.}). These combinations
of $G$ and $G^c/G$ currents, have the appropriate complex structure
required to be related to the currents following from the ($\chi$,
$\rho$) ghost system. Therefore, they enter naturally into the BRST
transformations.

  The BRST currents are $J_z^B = \chb \, (\CJ_z + \salf J_z^{gh})$ for
the left movers and $J_{\bar{z}}^B = \chi \, (\bar{\CJ}_{\bar{z}} +
\salf\bar{J}_{\bar{z}}^{gh})$ for the right movers. They give rise to
the following BRST transformation laws  \eqn\brl{ \delta
\CJ_{\bar{z}}= i\,\pzb\ch + i [\chi,\CJ_{\bar{z}}], \quad \delta
{\bar{\CJ}_z}= i\,\pz\chb + i[\chb,\bar{\CJ}_z]} \eqn\brm{ \delta \rh
= \CJ_z + J_z^{gh} \equiv \CJ_z^{tot}, \quad \delta \rb =
\bar{\CJ}_{\bar{z}}+ \bar{J}_{\bar{z}}^{gh} \equiv
\bar{\CJ}_{\bar{z}}^{tot}\,\,\, {\rm and}\,\,\, \delta \chi =\delta
\chb = 0.} By the $\ch$ equations of motion, $\delta
\bar{\CJ}^{tot}_{\bar{z}} = 0$ and $\delta \CJ^{tot}_{{z}}=  0$ and
$\delta^2=0$ ensues.

  $T_{zz}$ is expressed in the Sugawara form in terms of the complex
currents $\CJ$ and $\bar{\CJ}$. This is a consequence of the
particular values of $k$ and $\bar{k}$, the Kac-Moody levels, in the
two parts of the bosonic Lagrangian. The $hh^*$ part looks formally
like an ordinary $G$ Kac-Moody at level $-\bar{k}= -k-2c_G$. This sets
$\CN_i$, the normalization of the bilinear $i_z\,i_z$ in $T_{zz}$, to
be \eqn\njnj{\CN_i = {1 \over -\bar{k}+c_G} =- {1 \over k+c_G}= -
\CN_j,} where $\CN_j$ is the Sugawara normalization of $j_zj_z$.
Consequently \eqn\tanb{T_{zz}~=~\CN_j \,\TR\,(\CJ_z \, \bar{\CJ}_z)
{}~+~ \TR\,(\rh\,\pz\chb) ~=~\delta \bigl(G_{zz}\bigr)~=~ \delta \bigl(
\TR\,(\rh\,\bar{\CJ}_z) \bigr).}

  \def\pum{&\!\!\!\!\!}

  $\rh \,\bar{\CJ}_z$ is actually $b_{zz}$, the Virasoro anti-\-ghost.
This is an  example of the Super-\-Sugawara construction giving the
supersymmetric partner  of $T_{zz}$ as the product of the current and
its anti-\-ghost. Since the Virasoro algebra is in the enveloping
algebra of $G_k$, its anti-\-ghost is a composite operator as well. A
similar understanding of the Virasoro ghost $c^{z}$ is still missing
in this picture (understanding $c^z$ could be helpful in coupling
$G/G$ to gravity). Since both $\CJ_z^{tot}$ and $T_{zz}$ are BRST
anti-\-commutators, they have zero central charges $k^{{tot}}=0$ and
$c^{{tot}}=0$, which can also be explicitly seen from \eqn\cto
{\matrix{k^{{tot}}\pum= k \Bigl[G_k\Bigr] \pum+ \,k\Bigl[{G^c/G}_{
(k+2c_G)}\Bigr] \pum+ \,k^{gh}  \pum=\pum k \pum-\pum(k+2c_G)
\pum+\pum 2c_G\pum=\, 0\pum\cr  {\rm\!\!\!\!\!\!\!\! and }\cr c^{{
tot}}\pum=\, c\Bigl[G_k\Bigr] \pum+\, c\Bigl[{G^c/G }_{(k+2c_G)
}\Bigr]\pum+ \,c^{gh} \pum=\pum { \displaystyle{k \, d_G \over k +
c_G}} \pum+\pum {\displaystyle{(k +2c_G)\, d_G \over k + c_G}} \pum -
\pum 2 d_G\pum=\,0\pum.\cr} } $d_G$ is the dimension of $G$ and also
the number of the $(\rho, \, \chi)$ ghost-pairs. Their zero modes,
$L_0$ and $\CJ^{u\,\,tot}_0$ (the index $u$ runs over the Cartan
sub-algebra of $G$), are by themselves BRST anti-$ $\-commutators and
therefore, annihilate the states in the $Q$ cohomology.\foot{The
general argument is the following \ref\wsb{E. Witten, Nucl.Phys. {\bf
B220} (1982) 253.}: if $X=[Q,\phi]_+$ is some $\ng=0$ hermitian
operator ($\phi$ is consequently $\ng=-1$ operator) then $X$ commutes
with the BRST operator $Q$ and we can diagonalize $X$ while taking the
$Q$ cohomology. Let $\ket x$ be  an $X$ eigenstate with the eigenvalue
$x$. If $\ket {x\!}$ is in the $Q$ cohomology, $Q\ket x=0$ is
required. Then $x\ket x = X\ket x = Q \rho \ket x$. Thus, $\ket x$ is
in the image of $Q$, unless $x=0$.} This provides an important guide
for working out this cohomology in the next section.

  Another remark is related to the Sugawara normalization factors in
\njnj. $N_j^{-1}=k+c_g$ has an additional meaning as the periodicity
of the $G_k$ conformal blocks (and their fusion) on the weight lattice
of $G$ \ref\fs{M. Spiegelglas, Phys.Lett.{\bf 245B} (1990) 169 and M.
Walton, Nucl.Phys. {\bf B340} (1990) 777.}. We notice that
$N_i^{-1}=-(k+c_g)$ in the $(G^c/G)_{ k+ 2 c_g}$ part. There is no
apparent meaning to conformal blocks in this part, as we have noticed
it is pretty much structureless. However, it has the formal appearance
of $G_{- k- 2 c_g}$ and fusion may well have meaning for it. Although
a lot have to be understood about this formal analogy, it is
reassuring to notice that the periodicities match.

\newsec {The Spectrum of $G/G$.}
\def\i{l}

  We have formulated $G/G$ theories via cohomology in a complex BRST
construction associated with the $G_k$ constraints. Imposing the
constraints reduces substantially the large number of states typical
to the $G_k$ WZW field theory. We have seen that a topological theory
emerges, with no propagating fields, which describes merely global
degrees of freedom. Quantum mechanically we have a finite spectrum (or
discretely infinite spectrum, as in the $SU(2)/\IR$ case we mentioned
in the previous section) since local fields, the usual cause for the
enormous Hilbert space of field theory, are missing in $G/G$. The
excited states of the $G_k$ theory disappear from the $G_k/G_k$
spectrum via the mechanism of Complex BRST cohomology. This gets rid
of states in complex pairs\foot{The ``split-\-$ $a-\-pair" mechanism
implies that two non-\-orthogonal zero norm states, schematically
called $\ket{{\rm a}_{\pm}}$, result from a negative norm state
$\ket{{\bf a}_0}$. $\ket{{\rm a}_+}$ is excluded by $Q\ket{{\rm
a}_+}\not= 0$. $\ket{{\rm a}_-}$ is then excluded from the cohomology,
being an image of $Q$. The pair $\ket{{\rm a}_{\pm}}$ can be viewed as
a complex pair.} \huch\ and is actually a quartet mechanism\foot{The
quartet mechanism implies that the a``splitted-\-pair" with ghost
number $\ng$, combines with two other excitations of ghost numbers
$\ng \pm 1$ to form a twisted $N=2$ multiplet, which in turn
disappears from the physical spectrum.}  \kaog$\, $\karb. CBRST was
described as twisted $N=2$ supersymmetry  in section 2. The
negative-\-norm decoupling mechanism provides,  therefore, another
argument for the complex $G^c$ formulation,  since it always involves
complex pairs. In YM theories and in string theory the non light-cone
bosonic degrees of freedom (those which eventually disappear via the
BRST cohomology) always form complex pairs  \huch$\, $\ref\prsu{M.
Spiegelglas, {Class.Quan.Grav.} {\bf 5} (1988) L59.}. CBRST (or
twisted $N=2$ supersymmetry) is thus a well-known mechanism which we
apply here for the case of current algebras.

\subsec {The Physical States.}

  We are going \me now to study the physical spectrum of the $G/G$
model by calculating the characters, which lead to the torus partition
function. We are going to see that the physical spectrum, reflected in
the partition function, consists of a finite number of states. We also
wish to study the BRST cohomology. We will find it to be much bigger
than the physical spectrum. The cohomology space contains infinitely
many states, which will be argued to be spread over all possible
values of the ghost number $\ng$. The physical spectrum on the other
hand, is restricted to the $\ng=0$ sector of this cohomology space
which is finite. All these states are zero eigenstates of the total
$L_0$, \mei as follows from  $L_0=[Q,b_0]_+$ and the arguments in the
last section. They are also zero eigenstates of the total current
$\CJ_0^{3 \,tot} =[Q,\rho^3_0]_+$ and thus singlets of total $G$.
Although the whole cohomology  exceeds the physical spectrum, we will
find its calculation challenging. Moreover, the results have
interesting physical implications and show an intriguing analogy with
the Liouville theory of two dimensional gravity coupled to matter.
Both the physical spectrum and the cohomology follows from the
calculation of the $G/G$ characters which we obtain following the work
of Gaw{\c{e}}dzki and Kupiainen \gak.

  In \gak\ the torus partition function was calculated for a general
coset model $G/H$, starting with the gauged WZW model and trading the
gauge field $\az$ for a complex gauge transformation $h$ (see the
previous section for more details). We will restrict our review of
this calculation to the simpler $G/G$ case ($H/H$ is however, a
crucial step towards $G/H$ \gak). We follow the framework presented in
the previous section. An essential difference is that we have so far
discussed the case of a topologically trivial world-sheet $\WS$. We
will indicate how to modify those arguments to the case of a torus,
which allows for non-trivial holonomies along the two basic homology
cycles. These holonomies can be viewed as Aharonov-Bohm fluxes running
inside (and outside) the torus. They prevent the gauge configuration
$\az$ from being written as a complex gauge transform of a zero gauge
field. In order to perform the functional integration, the holonomies
will be expressed by writing $\az$ as the complex gauge transform of a
non-trivial standard gauge configuration $\cz$, \ie\ $\az = i\, h^{-1}
\pz h + \cz$, $\azb = -i\,h^{*} \pzb {h^{*}}^{-1} + \czb$. We will
have to include integration over the holonomies (parametrizing the
$\cz$'s) as a part of the functional integration. To be more specific,
let us recall from the previous section that $h(z)$ can account for
all the local features of the gauge configurations (like curvature). A
map from the fundamental group $\pi_1(\WS)$ to $G$ accounts for the
holonomies \atbt. This map is specified through a flat gauge
configuration (which fixes our standard for $\cz$) or equivalently, by
two commuting elements of $G$ to express the two holonomies.
Therefore, the $d[A]$ functional integral turns into the $d[h]$
functional integral plus an ordinary $du$ integral over a complex
parameter residing in the Cartan subalgebra of $G$, which accounts for
the two holonomies \gak.

  The $G/G$ torus partition function $Z_{G/G}$ is expressed in \gak\
as \eqn\zgg{ Z_{G/G}\,= \,C \tau_2^{-r} \int Z^g(\tau,u) \,\, Z^{hh^*}
(\tau,u) \,\, F(\tau,u) \,\,du\,,} where $du$ is the measure over the
moduli space of flat $G$ connections on the torus and $r$ is the rank
of $G$. $Z^g$ is the (torus) partition function for the WZW model
based on $G_k$. It is given by the sum over the $G_k$ conformal
blocks: \eqn\zg{ Z^g(\tau,u)\, =\, (q \bar{q})^{-c/24}\, \sum_{
\lambda_L, \lambda_R} N_{\lambda_L,\lambda_R} \chi_{k,{ \lambda_L}}
(\tau,u)\,\, \overline{\chi_{k, {\lambda_R}} (\tau,u)}.} In \zg\ $q=e
^{2 \pi i \tau}$, $\tau$ being the modular parameter of the torus.
$c=c[G_k]$ is the Virasoro central charge for $G_k$, given by \cto.
$\lambda_L$ and $\lambda_R$ are $G_k$ highest weights and each term in
the sum \zg, is a product of a right and a left $G_k$ Kac-Weyl
characters written as \kac\gepw \eqn\mm {\chi_{k,\lambda} (\tau,u) ~
=~ {M_{k,\lambda} (\tau,u) \over M_{0,0} (\tau,u)}.} $N_{\lambda_L,
\lambda_R}$ is the positive and symmetric ``mass matrix'' for the WZW
model which is  $N_{ \lambda_L, \lambda_R} =\delta_{\lambda_L,
\lambda_R}$ for the diagonal modular invariants of a simply-connected
group $G$ (we will restrict ourselves to those cases). $Z^{hh^*}
(\tau,u)$ in \zgg\ is the contribution of $h\in G^c/G$, at level
$k+2c_G$, to the partition function ($c_G$ is the dual Coxeter number
of $G$). This was calculated in \gak, using the iterated Gaussian path
integration technique.  $G^c/G$ is topologically trivial with only one
conformal block. \eqn\zh{Z^{hh^*} (\tau,u)\, \propto\, {\vert M_{0,0}
(\tau,u) \vert} ^{-2}} $M_{0,0}(\tau,u)^{-1}$ thus gives the level
$k+2c_G$ character of $G^c/G$ (which can also be viewed as the level
$-k-2c_G$ character of $G$) up to normalization. $F(\tau,u)$ in \zgg\
is the Fadeev-Popov ghost determinant which also factorizes
holomorphically. Conformal blocks appear only in the \mei $G_k$
sector,\foot{It would be nice to relate the appearance of more than
one conformal block in the $G_k$ sector, as well as other algebraic
features of $G_k$ (like null states), resulting in a non-trivial
numerator, to the topology which distinguishes it from the
$(G^c/G)_{\bar{k}}$ sector.} \me which is also the only source for a
numerator in $Z_{G/G}$ (the Kac-Weyl numerator $M_{k,\lambda}$). This
is due to the presence of algebraic relations corresponding to null
states only in the $G_k$ sector of the $G_k/G_k$ theory. $G^c/G$ is
free of algebraic relations and their manifestation in the form of
null states and thus results in a single Verma module (this is for the
case where $k$ is an integer, restricting ourselves to integrable
representations. A richer structure is revealed \ref\mff{ F.G.
Malikov, B.L. Feigin and D.B. Fuks, Funkt.Anal.\-$ $Prilozh. {\bf 20}
(1989) 25.} when these restrictions are lifted). Therefore, when in a
short while, we will study the cohomology, the important algebraic
arena will still be $G_k$ itself and will employ mainly representation
theory for $G_k$.

  We learn from $Z_{G/G}$ about the spectrum. More detailed
information about it, follows from a careful examination of the part
holomorphic in $q$, which we call the $G_k/G_k$ character.
$\Theta={\rm Re }\,u $ will be sufficient for our study of the
spectrum and analyzing it in terms of $\bar{G}$ multiplets (the global
$\bar{G}$ subgroup is generated by $j_0^a$, the zero modes of the
$G_k$ currents). This actually turns the torus into a cylinder, having
one \mei holonomy $\Theta$ around it and also one representation
$\lambda$ along it. Following \gak\ the  $G_k/G_k$ character is given
by multiplying $\chi_{k,\lambda} (\tau,\Theta)$ (the $G_k$ character),
$M_{0,0}(\tau,\Theta)^{-1}$ (for $G^c/G$) and $F(\tau,\Theta)$ (the
ghost contribution). Apart from a power of $q$ normalizing the
character, $G^c/G$ contributes $M_{0,0} (\tau,\Theta) ^{-1}$.
$F(\tau,\Theta)=M_{0,0} (\tau,\Theta)^{2}$ (again, up to
normalization), is the Fadeev-\-Popov determinant of the differential
operator $\pz$ in the $\rb \,\pz \ch$ term of \stol. Therefore, the
$G_k/G_k$ character is given by $M_{{k,\lambda}} (\tau,\Theta)$, the
numerator of $\chi_{ k,\lambda}$, the corresponding $G_k$ character.

  \mei This $G/G$ character, given by the Kac-Weyl numerator
$M_{{k,\lambda}} (\tau,\Theta)$, has an interpretation as a Kac-Weyl
``boundary" (the numerator of a character, which can be viewed as a
``reduction" of a representation which is reconstructed by \mei
``filling"),  introduced in \bndr. $\chi_{ k,\lambda}$, the original
character of $G_k$, decomposes into various $\bar{G}$ multiplets, with
each power of $q$ expressed as a sum of $\bar{G}$ representations.
Similarly, the expansion of the numerator ${M_{k,\lambda} (\tau,u)}$
contains the ``boundaries" of $\bar{G}$ representations. These
$\bar{G}$ ``boundaries" are numerators of the Weyl character formula
(\mei expressed as exponentials in $\Theta$ summed over the Weyl group
of $\bar{G}$). The Kac-Weyl ``boundary", given by the Kac-Weyl
numerator, amounts to a more radical reduction since we have imposed
the whole $G_k$ algebra in addition to $\bar{G}$. The states accounted
by this $G/G$ character contain neither descendant states present in
the original $G_k$ representation, nor excitations of ghosts and
$G^c/G$. The cohomology described by this character contains the
primary states. It does not contain any of their excitations, due to
``denominator cancellation". The other states which do appear in the
$G_k/G_k$ cohomology are associated with the null states of the
original $G_k$ theory, ``dressed'' by the ghosts and $G^c/G$, as is
evident from the $q$ expansion of $M_{{k,\lambda}} (\tau,\Theta)$ (the
``ghost dressing" accounts for the alternating signs in the $q$
expansion). We will further elaborate on these extra states through
this section.

  A notable similarity exists between the Kac-Weyl ``denominator
cancellation" in $G/G$ and the mechanism canceling oscillator
excitations in characters of $c=1$ matter systems coupled to two
dimensional Liouville gravity \bek\ ($c<1$ systems also follow along
similar lines). In the Liouville case too, the contributions from the
oscillator excitations to the character cancel between the three
constituent of the theory \ie\ the $c=1$ matter, the Liouville field
and the $(b,c)$ ghosts. The structural analogy to $G/G$ which we have
pointed out in the introduction, is strengthened when realizing that
``denominator cancellation" is responsible for the disappearance of
excitations in the Liouville case as well. For a $c=1$ matter system
coupled to Liouville let $\eta(q)$ denote $\prod_{n=1 }^\infty
(1-q^n)$. $\eta(q)$ serves as the denominator in the character of the
matter system, where it accounts for the excitations due to
oscillators. Combined with another $\eta(q)^{-1}$ factor from the
Liouville character it cancels against the ghost character,
$\eta(q)^2$. Thus, once the contribution due to the oscillator
excitations is canceled, the stage is left clear for the topological
properties of the matter system manifested in the $c=1$ numerator,
such as winding. Note that the analogy we are pursuing here between
the $G/G$ theory and the $c=1$ matter coupled to gravity goes one step
further. Both the Liouville and the $G^c/G$ sectors show no null
states and thus contribute simple factors ($\eta(q)^ {-1}$ and
$M_{0,0} (\tau,\Theta) ^{-1}$ respectively) to the total character.

  Now, we would like to employ the character found above to extract
the spectrum of the $G/G$ theory. For the sake of simplicity we will
concentrate in our study on the $SU(2)_k$ case, although the
conclusions will be general. $\Theta$ is then simply an angle $\theta$
and \mei integrable multiplets (or conformal blocks) are characterized
by the spin $j$ of the global $\overline{SU(2)}$. The spectrum can be
read from the character in two ways. The first one is by calculating
the partition function from the characters \gak. This gives the trace
over the physical space, or its dimension. Since the $G/G$ characters
are orthonormal in the $du$ measure $\Bigl(\int du\, M_{{k,j}}
(\tau,u)\, \overline{M_{{k,j^{'}}}(\tau,u)}= \delta_{jj^{'}}$ where
$j$ and $j^{'}$ denote multiplets of $SU(2)_k\Bigr)$, $Z_{SU(2)_k
/SU(2)_k}=k+1$. This is the number of conformal blocks of $SU(2)_k$
confirming that there is {\sl one physical state per block}.

\subsec {More Physical States--The BRST Cohomology.}

  We can, however, get more detailed information about the CBRST
cohomology by a closer look at the $G/G$ character, which was found
above to be the Kac-Weyl numerator $M_{k,j} (\tau,\theta)$. We will
see that $M_{k,j} (\tau,\theta)$ serves as an index, or the Euler
number, for the current-algebra cohomology in our CBRST complex \huch.
This follows from realizing that the $SU(2)_k$ ghost contribution
$F(\tau,\theta)$ is actually calculated with twisted boundary
conditions in the ``time direction" (a common practice in
supersymmetric theories). Thus, the trace expressed in $F(\tau,
\theta)$ contains $(-1)^{\ng}$,  as implied in the product \eqn\ft{
{F(\tau, \theta)= M_{0,0} (\tau,\theta )^{2}= {\prod_{n=1} ^{\infty}
\Bigl( (1-q^n) (1-q^n e^{i \theta}) (1-q^n e^{-i \theta}) \Bigr)^2}}.}
$\Bigl( (-1)^{\ng}$ is responsible for the signs of $q^n$ in \ft. \me
It amounts to the change $q^n \rightarrow -q^n$, transforming the
ordinary fermionic partition function into \ft.$\Bigr)$ The index
$M_{k,j} (\tau,\theta)$ is particularly useful as a formal power
series in $q$ (and  $e^{i\theta}$); with the term $a_\n0 \, q^\n0$
giving rise to $a_\n0$, the Euler number for the subspace of states of
{\sl excitation level} $\n0$. The formal power series allows us to
cover all the excitation levels $\n0$ at once. We note that the $G/G$
physical state $\ket{m_{k,j}}$, corresponding to the $j$'s conformal
block, is the  highest $\overline{SU(2)}$ weight component of the
$\n0=0$ excitation level in the cohomology. It is represented by the
term with the highest  $e^{i\theta}$ power and the lowest $q$ power in
$M_{k,j} (\tau,\theta)$. It has the physical value for the ghost
number \ie\ $\ng=0$ (or a zero degree form, if we use the de-Rahm
analogy \grsw\ ). This is the $G_k$ highest weight state discussed in
section 3 and counted by $Z_{G_k/G_k}$. We will see how \mei a richer
$G_k/G_k$ cohomology is unfolding via the index $M_{k,j}
(\tau,\theta)$.

  For $SU(2)_k/SU(2)_k$ the character for the $j$'s multiplet is
\eqn\mjk{M_{k,j}(\tau,\theta) \, =\,\sum_{\ell= - \infty}^{\infty}
q^{(k+2)\Bigl( \ell+{2j+1 \over 2(k+2)} \Bigr) ^2} \sin \Bigl( (k+2)
\ell + {2 j + 1 \over 2} \Bigr)\theta.} It combines ``boundaries" of
spin $(k+2) \ell + j$ $\overline{SU(2)}$ multiplets which result from
$\ell>0$ and negative ``boundaries'' of spin $(k+2) \ell + j-2$ from
$\ell<0$. We are trying to read, as much as possible, about the
cohomology by interpreting $M_{k,j} (\tau,\theta)$ as the index.
Clearly, the mere fact that $M_{k,j}$ is given as an infinite power
series tells us that each physical highest weight state gives rise to
an infinite number of states in the cohomology.

  We have denoted the excitation  level (\ie\ the power of $q$) by
$\n0$. $\n0$ is the contribution to $L_0$ from the non-\-zero mode
excitations of the currents and the ghost fields: \eqn\nfs{\n0=
\sum_{m \not=0} \Bigl(\CN_j\, :j_m^a \,j_{-m}^a: \,+\,\CN_i\,
:i_m^a\,i_{-m}^a \,+\, \rho_m^a \chi_{-m}^a \Bigr).} $\n0$ thus
includes contributions from the $m \not= 0$ modes of $j_{m}^a$ (the
excitations of the $SU(2)$ currents), $i_{m}^a$ (the excitations of
the $\bigl (SL(2,\IC) /SU(2) \bigr)_{k + 4}$ currents) and the ghosts
($ \rho_m^a,\, \chi_{-m}^a$). $\n0$ is devoid of of the zero mode
contributions $j^a_0$, $i^a_0$, $\rho^a_0$ and $\chi^a_0$. Their
inclusion renders $L_0=0$ and $\CJ_0^{3 \,tot}=0$ on the $G/G$
cohomology. $L_0=0$ results here in a close analogy to the way it
results on the Virasoro algebra cohomology (as employed for the
no-\-ghost theorem  \huch$\, $) in the critical strings case. For
critical strings, the total $L_0$ receives contributions from the
string excitations $\n0_{str}$ and from the zero modes. $L_0=0$ on the
string physical states results from the cancellation of these two
contributions.\foot{ In the bosonic string, for instance,
$L_0=-(p^2+1)+\n0_{str}=0$ is the mass shell condition. $p^2$ which is
associated with the bosonic zero modes (and 1 which can be traced to
the ghost zero modes) cancel against the contribution of the
excitations $\n0_{str}$.} $\n0_{str}$ is the excitation level in the
index found in \huch, like $\n0$ in the $G/G$ case.

  The group structure in $G/G$ yields more information which allows us
to arrange the states appearing in the $q$ expansion of the index,
into multiplets of $\bar{G}$. This information is provided by the
$e^{i \theta J_3}$ factor included in the index. $J_3$ is the sum of
the zero modes of two out of the three contributions to the total
current,\foot{ $\CJ_0^{3 \,tot}=0$, follows from the cancellation of
$J_3$ against $i^3_0$. $i^3_0$ generates the Cartan subalgebra of the
$SL(2,\IC) /SU(2)$ part. Its analogy to $p$, which renders $L_0=0$ in
the bosonic string case will become apparent soon. We will see in the
following $SU(2)/SU(2)$ examples that $i^3_0=-J^3$ also ensures
$L_0=0$, thus providing an independent check for our construction. }
namely, $J_3 = j^3_0 + J^{3\,gh}_0$ with $j^3_0$ due to $SU(2)$ and
$J^{3\,gh}_0$ due to the ghosts. Now we are in the position to regard
$M_{k,j} (\tau,\theta)$ as the index ${\rm Tr}\,(-1)^{\ng} q^{\n0}
e^{i \theta J_3}$ and read the cohomology from it \ref\ber{The
cohomology was considered independently, along similar lines by M.
Bershadsky.} \mei Here we will outline, how this interpretation, which
we offer for the power expansion of $M_{k,j} (\tau, \theta)$,
correctly identifies the \mei  $G/G$ cohomology, with the details
supplied in \ref\cmg{M. Spiegelglas and S. Yankielowicz, {\it
$G/G$--Current Algebra Cohomology from the Kac-Weyl Numerator} TAUP-$
$\-Technion preprint, in preparation.}. \me We will also give a
heuristic argument, taking stock of the algebraic relations available
for the $G/G$ models, that we have identified the {\bf whole}
cohomology. It is reassuring that this argument seems to agree with
the rigorous result for this semi-infinite cohomology \ref\efr{E.
Frenkel, Private communication.}.

  The interpretation \me of the index $M_{k,j} (\tau, \theta)={\rm Tr}
\,(-1)^{ \ng} q^{\n0} e^{i \theta J_3}$ as the Euler number allows us
to study the states in the cohomology. The values of $\n0$ gives the
excitation level, $J_3$ gives the ``angular momentum" and $\ng$ gives
the ghost number (only its parity, strictly speaking) of these states.
Usually, the Euler index does not determine the states and due to
cancellations it only indicates few of the states in the cohomology (a
very small fraction in most cases, which usually prevents us from
deducing the Betti numbers from the Euler characteristics). Here,
there seems to be an exception to this. We will see that a lot can be
learned about the cohomology from the index. It is clear from \mjk
that {\sl there are infinitely many states in the cohomology for each
conformal block $j$}, at least one state per term in the sum over
$\ell$, since all the contributions to the index are due to states in
the cohomology. In the next subsection we will use this interpretation
to delve deeper into the structure of the whole cohomology. We will
suggest an iterative procedure for the actual construction of the
states. Although we do not carry the full construction, there will be
enough indications to provide compelling arguments supporting the
conclusion that for each conformal block there is one state in the
cohomology for every possible $\ng$ value.

\subsec {States at all Levels--Cohomology from Algebraic Relations.}

  We start our construction of the CBRST cohomology by using
information which is encoded in the index $M_{k,j} (\tau, \theta)$\ $=
{\rm Tr} \,(-1)^{ \ng} q^{\n0} e^{i \theta J_3}$ to guess the
``leading" states $\ket{f^j_\ell}$  ($j$ is the $SU(2)_k$
representation while $\ell$ refers to the $\ell$'th term in the
expansion of $M_{k,j}$ in \mjk) The ``leading" states, which we
define, are the ``leading components" (LCs) of the actual states $\ket
{\psi^j_\ell}$ in the cohomology. The LCs constitute the starting
points for an iterative procedure for the construction of $\ket{
\psi^j_\ell}$. ``Leading" states are constructed by hitting the ground
state $\ket{m_{k,j}}$ (with $j^3_0=j$) by $o^j_\ell$, a minimal degree
monomial in the currents and ghost excitations. Recall that for a
physical highest weight state $\ket{m_{k,j}}$, encoded in the $\ell=0$
term, $J^3$ is saturated by $j$. $\CJ^{tot}_3=0$ determines the
$i^3_0$ eigenvalue to be $-j-1$ (the shift by one unit, is due to the
ghost zero mode involved in $\ket{m_{k,j}}$).  For the higher terms,
$o^j_\ell$ is constructed to be of minimal degree saturating the
leftover  $J_3-j$ and the level $\n0$, as read from the index. This
choice allows, therefore, only the negative modes of the $SU(2)_k$
currents $j^+_{-m}$, ghosts $\chi ^+_{-m}$ and anti-ghosts $\rho
^+_{-m}$ to appear in $o^j_\ell$. It excludes zero modes and
$i^a_{-m}$ components (those would increase $\n0$ with no effect on
$J_3$). Moreover, we would like $o^j_\ell$, to be iterative \ie\ to
have $o^j_{\ell}$ as a factor in $o^j_{\ell+1}$.  To account for the
alternating parity of the ghost number (indicated by $(-1)^{\ng}$ in
the index) we choose to have in   $o^j_{\ell}$ only anti-ghost modes
for the $\ng<0$ states and only ghost modes for the ``dual" $\ng>0$
states. This implies that  that one ``leading" state appears for each
value of $\ng$.

  There is a lot of arbitrariness in our ``educated guess" for the
``leading'' states. We do not have an  a-priori justification for all
the assumptions we made in this construction. The true check would be
provided by producing the states $\ket{\psi ^j_\ell}$ in the
cohomology via an iterative procedure, starting from these LCs. This
task of building the states $\ket{\psi ^j_\ell}$ as polynomials in
excitations (\ie\ as sum over states generated by non-minimal
monomials acting on various ground states, characterized by different
eigenvalues of the zero modes) and checking that they are in the
cohomology, is pretty tedious and is therefore, deferred to \cmg.
Here, we will just discuss the crucial point of using the {\sl
algebraic relations,} (that shape the $j$ representation of $SU(2)_k$
by supplying null states for it) to verify that the $\ket{\psi
^j_\ell}$ states in the cohomology could be constructed iteratively,
starting from the guessed ``leading" states $\ket{ f^j_\ell}$. We
observe that there is an algebraic relation corresponding to every
state $\ket{ f^j_\ell}$. This relation should be employed to check
that $Q$ annihilates $\ket{ \psi^j_\ell}$. Since we did not construct
$\ket{ \psi^j_\ell}$, we will only do the first stage of this
verification and examine $ Q\, \ket{ f^j_\ell}$. We will see that the
LC (``leading component") of $ Q\, \ket{ f^j_\ell}$ vanishes, due to
the algebraic relation. There is a ``non-leading" part to $ Q\, \ket{
f^j_\ell}$ which is canceled by the ``non-leading" terms of $\ket{
\psi^j_\ell}$. Those can be constructed iteratively \me with the
details deferred to \cmg. The check that $\ket{ \psi^j_\ell}$ is not
in the image of $Q$ is easy. We will see in a while, that there could
be no state whose $Q$ image is $\ket{ f^j_\ell}$. Thus also $\ket{
\psi^j_\ell}$ can not occur in the image of $Q$. For this argument,
choosing $o^j_\ell$ to contain only anti-ghost modes is crucial. We
should comment, that for the ``dual" $\ng>0$ states with only ghosts
in their  corresponding $o^j_\ell$'s, this picture reverses. While by
construction the $\ng>0$ states are annihilated by $Q$, the algebraic
relations are needed to guarantee that these states are not in the $Q$
image.

  The fact that algebraic relations and the associated null states are
present in the representation of $SU(2)_k$ is helpful in two ways. It
verifies that the ``leading" states are in the cohomology, in the
``leading" sense (\ie\ up to ``non-leading corrections it is in the
kernel of $Q$ and not in its image), thus offering the first stage of
an iterative process of corrections. This process results in the state
$\ket{ \psi^j_\ell}$ belonging to the cohomology. This state has
$\ket{ f^j_\ell}$ as its LC and the appropriate corrections involve
combinations of $i_m^a$ and the zero modes. The lack of relations in
the $SL(2,\IC)/SU(2)$ sector seems to allow this iterative process to
go unstopped, canceling ``non-leading" contributions in stages. This
brings us to the observation that the cohomology is a manifestation of
the algebraic relations in the $j$ multiplet of $SU(2)_k$. This
observation implies that having \me utilized all the algebraic
relations, we have found the {\bf whole} cohomology from the Euler
index. From our construction it is evident that we will not be able to
find a state in the cohomology unless it is \me ensured by the
algebraic relations. The same holds for the ``dual" states (with
ghosts rather than anti-ghosts in their $o^j_\ell$ monomials). These
are trivially annihilated by $Q$. The algebraic relations are
essential in verifying that they are not in its image.

\def\rh{\rho} \def\ch{\chi}

  As our first simple example, we take the case of complex cohomology
for the Lie group SU(2). The Weyl numerator for the multiplet $j$ of
$SU(2)$ is  \eqn\wey{M_j(\theta) = \sin (j+\half)\theta= {1 \over 2i}
e^{i{(j+ \salf) \theta}}\Bigr( 1- e^{-i{(2 j+ 1 )\theta}} \Bigl).} We
are studying the complex generated by the $SU(2)$ generators $j^+, \,
j^- ,\,j^0\, , $ by their $SL(2,\IC) /SU(2)$ counterparts $i^+ ,i^-,\,
i^0\, , $ by the anti-\-commuting ghosts $\chi^+, \chi\,^-,\, \chi^0$
and the anti-\-ghosts $\rh^+,\, \rh^-,\, \rh^0$.

  Interpreting $M_j(\theta)$ as  ${\rm Tr}\,(-1)^{\ng} e^{i \theta
J_0}$ would lead us to the conclusion that the first term  in
$\bigl(1- e^{-i(2 j+ 1)\theta}\bigr)$ is due to the ground state
$\ket{-j, \, j}$, which is a lowest weight state of $j$ as well as a
highest weight state of $i$ ($\,j^0 \ket{a , \, b} = a \ket{a, \, b}$
and $i^0 \ket{a, \, b} = b \ket{a , \, b}$). The second term stands
for a state with ${\ng=-1}$. The ``leading" state is $\ket{\ng=-1}
\,=\, \rho^+ (j^+)^{2j} \,\ket{-j ,\, -(j+1)}$ in this case $\bigl($to
keep $\CJ^0_{\rm tot}=0$, the $i^0$ value changes by  $(2j+1)\bigr)$.
Verifying that $Q \vert \ng=-1>=0$, relies on $Q\ket{-j , \, j}=0$ and
on $[Q,\rho^+]{\strut_+}=j^+\, + \,i^+$. The LC of $Q\ket{\ng=-1}$ is
${(j^+ )}^{2j+1} \,\ket{-j , \, -(j+1)} =0$ (ignoring the $i^+$ from
$[Q,\rho^+]{\strut_+}$, which results in a ``non-leading"
contribution). The last equality is simply the defining relation for
the $j$'s  $SU(2)$ multiplet (defined via a lowest weight state
annihilated by ${(j^+)}^{2j+1}$). Denoting all the states in the
cohomology generically by $\ket{\psi^j_t}$ and its LC by
$\ket{f^j_t}$, we can now state our main result: {\sl The ``$G/G$
character" given by the Weyl numerator for the representation $j$,
encodes all the states $\ket{ \psi^j_t}$ in the cohomology. $\ket{
\psi^j_t}$ are constructed to satisfy $Q\,\ket{ \psi^j_t}=0$, in an
iterative process starting from the ``leading components" $\ket{
f^j_t}$. The ``leading" parts of $Q\,\ket{ \psi^j_t}$ are the
``leading" parts of $Q\,\ket{ f^j_t}$ and vanish due to the algebraic
relations satisfied in the representation $j$. Since $Q\,\ket{
\psi^j_t}=0$ have consumed all the algebraic relations, they
constitutes all the $G/G$ cohomology.}

  Returning to complex current algebra cohomology, we are going to see
a similar picture in the $SU(2)_2$ example. Let us write down the
monomials in currents and ghosts excitations $o^j_\ell$, for the LCs
of the states in the CBRST cohomology implied by $M_{2,j}
(\tau,\theta)$ \eqn\chrn{ {M_{2,0} (\tau,\theta) \,\propto\,
\sum_{\ell= - \infty}^{\infty} q^{\ell\,(4\ell+1)} \sin \,(\, 4 \ell +
{ 1 \over 2}\,)\,\theta}=} $$\matrix{{\hbox{\hfil$=$}} &{\bf 0}
&-&q^3\,{\bf 3}\, &+&q^5\,{\bf 4}\, &-& q^{14}\,{\bf 7}\,
&+&q^{18}\,{\bf 8}\, &-+\! &\ldots \cr \cr {\rm  LCs:} &&&\!\eta_1
\i_1^2 &&\!\eta_1\eta_2 \i_1^2 && \!\eta_1\eta_2\eta_3 \i_1^2 \i_3^2
&&\!\eta_1\eta_2\eta_3\eta_4 \i_1^2 \i_3^2& &\ldots\cr\cr }$$

  \smallskip For the sake of concise notation, we use $\eta_n$ to
denote $\rho^+_{-n}$, the anti-ghost creation operator\foot{
$\eta_n=\chi^+_{-n}$ is a dual choice producing states with positive
ghost number, which will be explored later in this secion.} and $\i_n$
to denote $j^+_{-n}$. Note that $\i_n$ is the LC of $[Q,\eta_{n}]
{\strut_+}$ (ignoring the $i^+_{-n}$ and the ghost parts which are
``non-leading"). \mei We use short hand notation for $\vert \sin (4
\ell + \salf)\theta \vert $ (\ie\ $ \sin (4 \vert \ell \vert +
\salf)\theta$ for $\ell \geq 0$ and $ \sin (4 \vert \ell \vert -
\salf)\theta$ for $\ell< 0$). These are viewed as Weyl numerators, or
``boundaries" of the $SU(2)$ multiplets, whose spin is $4 \vert \ell
\vert $ (for $\ell \geq 0$) and $4 \vert \ell \vert -1$ (for when
$\ell < 0$). Thus ${\bf 3}$, for instance, in the $\ell=-1$ term of
\chrn, denotes the term $\sin 3\salf\theta$ arising from the $SU(2)$
multiplet of spin 3. Decoding the cohomology from the $j=\half,1$
blocks:

  \eqn\chrh{{M_{2,\salf}(\tau,\theta)\,\propto\, \sum_{\ell= -
\infty}^{\infty} q^{\ell\,(4\ell+2)} \sin \,(\, 4 \ell + { 1
}\,)\,\theta}=} $$\matrix{{\hbox{\hfil$=$}}&{\!\!\!\balf }
&\!\!-&q^2\,{\bf 2 \balf }\, &\!\!\!+&q^6\,{\bf 4\balf}\, &\!\!\!-&
q^{12}\,{\bf 6\balf}\, &\!\!\!+&q^{20}\,{\bf 8\balf}\, &\!\!\!\!-+\!\!
&\!\!\!\ldots \cr \cr {\rm LCs:} &&&\!\!\!\eta_1 \i_1 &&\!\!\!
\eta_1\eta_2 \i_1 \i_2 && \!\!\!\eta_1\eta_2\eta_3 \i_1 \i_2 \i_3
&&\!\!\!\eta_1\eta_2\eta_3\eta_4 \i_1 \i_2 \i_3 \i_4&
&\!\!\!\ldots\cr\cr }$$

  \eqn\chro{{M_{2,1}(\tau,\theta)\,\propto\, \sum_{\ell= -
\infty}^{\infty} q^{\ell\,(4\ell+3)} \sin \,(\, 4 \ell + { 3 \over
2}\,)\,\theta}=} $$\matrix{{\hbox{\hfil$=$}}&{\bf 1} &-&q\,{\bf 2}\,
&+&q^7\,{\bf 5}\, &-& q^{10}\,{\bf 6}\, &+&q^{22}\,{\bf 9}\, &-+\!
&\ldots \cr \cr {\rm LCs:} &&&\!\eta_1  &&\!\eta_1\eta_2 \i_2^2 &&
\!\eta_1\eta_2\eta_3 \i_2^2 &&\!\eta_1\eta_2\eta_3\eta_4 \i_2^2
\i_4^2& &\ldots\cr\cr }$$

  Let us be more explicit about the states \mei which we have decoded
above; with $\ket{f^j_\ell}$ denoting the LC of the state\foot{\mei
Note that the choice we had for $J_3$ looks now more natural. Like
$\n0$ in the index ${\rm Tr}\,(-1)^{\ng} q^{\n0}  e^{i \theta J_3}$,
$J_3$ also counts the excitations of ghosts and currents in
${o^j_\ell}$ (apart from $j$, a $j^3_0$ contribution, which is
constant on a given $j$ multiplet). $i^3_0$ is purely due to zero
modes as we are going to discuss soon.} resulting from the $\ell$'th
term of the index $M_{2,j}$. We will demonstrate that the LC of $Q
\ket{f^j_\ell}$, which we denote by $\ket{g^j_\ell}$, vanishes
indicating that the whole state $\ket{ \psi^j_{\ell}}$ could be
constructed in the kernel of $Q$. $\ket{g^j_\ell}=0$ follows from
relations satisfied in the $j$ multiplet (highest weight
representation) of $SU(2)_2$. The verification that $\ket{\psi^
j_{\ell}}$ is in the kernel of $Q$, thus reduces into an algebraic
problem in $G_k$ representation theory (followed by an iterative
construction for $\ket {\psi^j_{\ell}}$, which is deferred to \cmg).
For $\ell=-1$ the analysis of $\ket{g^j_\ell}$ is simple and will be
presented here for the general $SU(2)_k$, as well.  From the $\ell=-1$
term we decode the LCs $\ket {f^j_{-1}}=\eta_1 \i_1^{k-2j} \ket j$,
where $ j_0^3 \ket j = j \ket j$. $\ket j$ denotes the ground state of
all non-\-zero modes and a highest weight state with respect to
$\overline{SU(2)}$ ($\ket j$ is also a $i_0^a$ highest weight. The
$i_0^3$ value will be discussed below). $\ket{g^j_{-1}}$, the
``leading part" of $Q\, \ket {f^j_{-1}} = Q\eta_1 \i_1^{k-2j} \ket j
$, is now $ (j^+_{-1})^{k+1-2j} \ket j =0$. This follows from a
general algebraic relation in the $j$'s representation of $SU(2)_k$,
asserting that $\ket j$ is also a lowest weight state in a multiplet
of $ {\bf S} \subset SU(2)_k$ as well.\foot{${\bf S}$ is the same
algebra used to find the integrable representations of $SU(2)_k$ in
\kac$\, $\gepw.} ${\bf S}$ is the $SU(2)$ subgroup generated by
$s^+=j^+_{-1}, s^-=j^- _{1}~ {\rm and}~ s^0=k/2-j^0_{0}$. Since $s^0
\ket j= (k/2-j) \ket j$, the ${\bf S}$ lowest weight state $\ket j$ is
annihilated exactly by $(j^+_{-1})^{k+1-2j}$.

  From the $\ell=1$ terms in the expansion of $M_{2,j}$, we find the
$\ng=-2$ states in the cohomology, $\ket {\psi_{1}^j}$ having their LC
equal to $\ket {f_{1}^j}= o_{1}^{\,j} \ket j$ with $o_{1}^0=\eta_1
\eta_2 \i_1^2$, $o_{1}^{{ \scriptscriptstyle 1/2}}=\eta_1 \eta_2 \i_1
\i_2$ and $o_{1}^1=\eta_1 \eta_2 \i_2^2$. We wish to show the
vanishing of $\ket {g_{1}^j}$, the ``leading" part of $Q\, \ket
{f_{1}^j}$. We consider here only the $SU(2)_2$ case \me utilizing the
fact that the $o_\ell^j$ are iteratively constructed. Notice that \mei
the state $\ket j$ is similar to to the state $\ket j$ in the
$\ell=-1$ case (differing only by the implicit value for $i_0^3$) and
$o_{1}^j$ still contains a $\i_1^{k-2j}$ factor. Concerned only with
the ``leading" $j_n^+$ part of $[Q,o_1^j] {\strut_+}$, no extra $\i_1$
can be generated (by the above ${\bf S}$ argument). Therefore, an
extra $\i_2$ is generated in the  anti-\-commu\-tation. This amounts
to a power of $\i_2$ which annihilates $\ket j$, by carrying it
``outside the $j$'s representation". In short, we have nailed down the
relations\foot{These relations describe the ``envelopes" of the $j$'s
multiplets of $SU(2)_2$. In terms of the ${\bf S}$ highest weight
state $\ket {s^0=1-j}=\i_1^{2-2j} \ket j$, the relations $\i_2^{2j+1}
\ket {s^0=1-j}=0$ in dicate the power of $\i_2$ ``missing" the $j$'s
multiplet. $[Q,o_1^j] {\strut_+}$ are ``trying" to increase $j_0^3$
too much. $j_0^3=j+3$ would have  been reached, which is just not
present at the $SU(2)_2$ excitation level of $\n0-3 =2j+4$.}
$\i_1^2\i_2 \ket{j=0} = \i_1\i_2^2 \ket{j=\salf} = \i_2^3 \ket{j=1} =
0$, which ensure $\ket {g_{1}^j} =0$. Similar relations appear for all
values of $\ell$ and ensure the vanishing of the ``leading" part of
$Q\, \ket {f_{\ell}^j}$. This in turn allows for an iterative
construction of the state $\ket {\psi_{\ell}^j}$ satisfying $Q\, \ket
{\psi_{\ell}^j}=0 $\cmg. $\ket {\psi_{\ell}^j}$ is the state in the
cohomology suggested by the $\ell$'th term in the expansion of
$M_{2,j}$.\foot{Checking that these states are not in the $Q$ image is
easy and as argued above, can be done for their LC. The $\ng=r$ LC
$\ket {\ng=r}$ contains the  $\eta_1, \ldots, \eta_r$ excitations and
no $\i_{t}$ for $r<t$. If  $\ket {\ng=r}= Q\, \ket {\ng=r-1}$, $\ket
{\ng=r-1}$ which is lower by one $\ng$ unit, must include an extra
anti-ghost creation operator $\eta_{u}$. For $u>r$, $Q\,\eta_{u}
\ldots$ should contain $i_{u>r}$ missing from $\ket {\ng=r}$. For $u
\leq r$ on the other hand, $\eta_u$ would render $\ket {\ng=r-1}=0$ by
the Pauli principle.}

  The argument presented here is by no means complete. \me It does
illustrate the availability of the algebraic relations needed to
ensure that $Q$ annihilates the states in the cohomology; states
enfolding through the index given by the Kac-Weyl numerator. The
precise way the $i$'s and the zero modes of the ghosts enter, will be
discussed appropriately in \cmg. Here we only wish to comment about
the value of $i^3_0$, which ensures that both $\CJ^{3 \,tot}$ and
$L_0$ vanish on the states that we have found. For this sake, we have
to assign the appropriate $i^3_0$ eigenvalue to each state (that we
have decoded from the $\ell$'th term) in \chrn,$\, $\chrh\ and \chro.
$i^3_0$ is taken to be $-J_3$, \ie\  $i^3_0=-(k+2) \ell - (j+ \salf
)$.\foot{\me The shift in  $i^3_0$ by half a unit, is accounted by the
choice of the ground state for the ghost zero modes.} In addition to
$\CJ^{3 \,tot}=0$, this value of $i^3_0$ renders the total $L_0$,
independent of both $\ell$ and the block $j$. The fact that choosing
the value of $i^3_0$, took care of both $\CJ^{3 \,tot}$ and $L_0$
provides an independent check of \me our evaluation of the cohomology
through the interpretation we offered for $M_{j,k}$. Moreover, $i^3_0$
turns out to be quantized, although a-priori it can be continuous. In
this, the $SU(2)/SU(2)$ cohomology is similar to the Virasoro discrete
states for $c=1$, which exhibit quantized momentum eigenvalues.
Another facet of the similarity between $SU(2)/SU(2)$ and the $c=1$
Virasoro discrete states was seen by the $SU(2)_1/ \IR$ interpretation
of the latter, offered in section 3. We should comment that the
$SU(2)_1/ SU(2)_1$ cohomology looks much like a refinement of
$SU(2)_1/\IR$.

  \me We have already commented that we can produce the LCs
$\ket{\phi ^j_{\ell}}$, for a similar family of states with $\ng>0$,
by taking $\eta_n$ to be the ghost creation operator $\chi^+_{-n}$ and
a suitable modification of $\i_n$. Checking that $\ket{\phi
^j_{\ell}}$ are annihilated by $Q$ proceeds by a ghost counting
argument, like in footnote {\global\advance\ftno by-1 {\the\ftno}
\global\advance\ftno by1}. The algebraic relations are then used to
show that $\ket{\phi ^j_{\ell}}$ are not in the image of $Q$. There
are thus, states in the cohomology for all values of the ghost number
$\ng$ and they are paired between $\ng$ and $-\ng$ \kaog$\, $\huch.

\subsec {``Dressed Null States"--Constituents of the Cohomology.}

\nref\klp{I.R. Klebanov and A.M. Polyakov, Mod.Phys.Lett.{\bf A6}
(1991) 3273.}

  We have discussed here the spectrum of the $G/G$ theory and
explained how the cohomology works to cancel the excited states. The
character for $G_k/G_k$ is the Kac-Weyl numerator $M_{k, \lambda}
(\tau,\Theta)$. It was also found to be the index ${\rm Tr} \,(-1)^{
\ng} q^{\n0}  e^{i \Theta \cdot J_h}$ ($\Theta \cdot J_h=\theta J_3$
for the $SU(2)$ case, or generally, a scalar product in weight space
of $G$) and as such has been used to identify the BRST cohomology. We
have established that even though, there is a single physical state
per conformal block (corresponding to $\lambda$, the highest weight
representation  of $G_k$), there are infinitely many states in the
$G/G$ CBRST cohomology which belong to this block. With the help of
some guesswork, we managed to argue that there is precisely one state
in the cohomology \me with a given $\ng$ value. These states are in
one to one correspondence with the various $\ell$ terms in the
expansion of $M_{k, \lambda} (\tau,\Theta)$. Their actual construction
follows through an iterative procedure starting from the ``leading"
components  for states with $\ng>0$ (and a dual construction for
states with $\ng<0$). The physical state is unique in the conformal
block and is found in the $\ng=0$ sector. The multitude of $\ng \not=
0$ states is nonetheless very intriguing in its close similarity to
the discrete states  \mei which constitutes an interesting part of
$c=1$ two dimensional gravity \wids$\, $\klp. \mei These extra
$\ng\not= 0$ states, encoded in the index $M_{k, \lambda}$, have their
origin at the null states of $G_k$, which are appropriately ``dressed"
by ghost and $(G^c/G)_{\bar{k}}$ excitations. Recall that the standard
r\^ole of the null states is to shape and bound the $G_k$ multiplets
(the same holds for Lie algebras). They correspond to the algebraic
relations defining the $\lambda$ representation of $G_k$. Once the
``denominator cancellation" takes place, the ``dressed" null states
reappear as genuine contributions to the $G_k/G_k$ character or to the
cohomology.

  The \mei cohomology, containing ``dressed null" states, is also
typical to the Liouville theory of two dimensional gravity. The matter
system is chosen to be a $c<1$ minimal model, or a $c=1$ RCFT. In the
gravitational case, the discrete states in the cohomology can be
traced to the matter null states $\ket {\delta,c}$ ``dressed" by the
Liouville state $\ket {1- \delta,26-c}$ and by ghost excitations. The
relation between the matter null state $\ket{\delta,c}$ (of anomalous
dimension $\delta$) and its Liouville ``dressing" $\ket {1- \delta,
26-c}$, is a manifestation of the ``duality" pointed out in \disc. We
expect a similar duality to hold between $G_k$ and $(G^c/G)_{k+2 c_G}$
(or $G_{-(k+2 c_G)}$)  \mff. We have noted the matching periodicities
in the previous section. We expect that this duality as well as the
analogy to \disc, can be drafted for the construction of
representatives for the cohomology classes and facilitate the tedious
iterative construction.

\newsec {Discussion and Outlook.}

  In the present paper we have constructed and investigated the $G/G$
topological theories. In our analysis we have employed both the CBRST
approach as well as a more direct approach based on the gauged WZW
model. The need for complexification was quite apparent in both
approaches. The topological theory based on $G_k$ contains a finite
number of physical states--$N_B$ which is the number of $G_k$
conformal blocks, as reflected in the calculation of the torus
partition function. These physical states are understood as the
highest weight states of the $G_k$ theory. They are related to  the
classical theory of flat $G$ gauge configurations\giii\ (quantized
with $1/k$ playing the r\^ole of Planck's constant). For $SU(2)_k$
$N_B=k+1$.

  This physical space of states is actually the zero ghost number
sector of the BRST cohomology discussed in section 4. For $\ng\not=0$
the rest of the cohomology can be viewed as extra discrete states,
encoded in the Kac-Weyl numerators which serve as the characters of
$G_k/G_k$. These states are one facet of the analogy between $G/G$ and
two dimensional gravity coupled to matter ($c=1$ matter was recently
receiving a great deal of attention \disc,\wids\ ). Another facet of
the analogy is the content of these theories. They both contain three
parts. The components of $G_k/G_k$ are $G_k$ WZW model, $(G^c/G)_{
\bar{k}}$ model (which is formally also the level $-(k+2\,c_g)$ $G$
WZW model) and spin $(1,\, 0)$ complex ghosts; whereas matter,
Liouville and $(2,\, -1)$ ghosts are the corresponding ingredients of
a two dimensional gravity system. To discuss this analogy further, we
have investigated the cohomology of $G/G$. An intriguing issue is the
correspondence between the $G_k$ representation and the particular
$(G^c/G)_{\bar{k}}$ representation which combines with it into a
$\ng\not=0$ state in the cohomology. We have explicitly analyzed here
the $G=SU(2)$ case. We would clearly like to verify this
correspondence for a general group $G$ and also to understand the
general principle behind it along with its counterpart in the case of
gravity. An approach to the Liouville theory which relies on current
algebra may be of help \kpz$\, $\nvil$\, $\blkd. An investigation,
along these lines, of the cohomology versus the physical space for
$G/G$ as well as for two dimensional gravity coupled to matter is
under its way.

  Amplitudes in the $G_k/G_k$ theory factorize in terms of the three
point functions. Those are in turn the $N_{ijl}$ \ie\ the fusion rules
for the $G_k$ WZW model. From this view point (employed in \giii\ to
calculate the $G/G$ amplitudes), $G/G$ theories are the two
dimensional counterparts of the three dimensional Chern-Simons
Gauge-Theories \csw. In CS the $N_{ijl}$ also give physical
amplitudes, or actually the partition functions in the $S^2\times S^1$
three dimensional topology. The $i$, $j$ and $l$ representations of
$G$ are located on three unlinked Wilson loops which run parallel to
the $S^1$. The $G/G$ amplitudes follow upon dimensional reduction,
ignoring the $S^1$.\mei This comes as no surprise since the CS
theories  are known to reduce to the fully gauged WZW model
\ref\mess{S. Elitzur, G. Moore, A. Schwimmer and N. Seiberg,
{Nucl.Phys.}{\bf B326} (1989) 108.}$\, $\witf.

  We still miss a genuine two dimensional CBRST understanding of the
amplitudes. This was clearly demonstrated in section 3, where we had
hard time studying the rational torus. Instead of constructing
$U(1)_{\nk}/ U(1)_{\nk}$ via CBRST cohomology,\foot{As a curiosity,
this attempted construction produced $SU(2)_1/\IR$, a WZW setting for
the Virasoro discrete states.} we had to resort to a description in
terms of gauge configurations and their holonomies.  Hopefully, a
thorough understanding of the $G_k$ current algebra cohomology and its
$\ng \not= 0$ richness, would create a promising alternative approach
for calculating the amplitudes. This hope is based on the analogy with
the superstring amplitudes which were properly calculated only after
the introduction of the ghost system and picture changing \ref\fms{D.
Friedan, E. Martinec and S. Shenker, Nucl.Phys. {\bf B271} (1986)
93.}. \me It is tempting to regard the states in the cohomology, with
various values of $\ng$ as the same state in different ``pictures". If
true, insertions in various ``pictures" should be used in the
calculation of an amplitude. Some evidence for this view is given by
the apparent equivalence of the different $\ng$ states in the
cohomology, where the difference is in the choice of the state being
called physical.

  The close analogy between $G/G$ theories and matter
coupled to two dimensional gravity in the Liouville approach, suggests
to look for a $2+1$ dimensional setting for the latter. Three
dimensional CS gravity is related to $SL(2,\IR)$ \ref\csg{E. Witten,
Nucl.Phys. {\bf B311} (1988) 46.} (\mei which tend to appear in the
Liouville \nvil\ and other two dimensional gravity contexts \kpz$\, $)
and other subgroups of $SL(2,\IC)$. The recent discovery of area
preserving diffeomorphisms in the study of $c=1$ matter systems
coupled to two dimensional gravity \disc$\, $\klp, is another
indication for a $2+1$ dimensional description. Higher dimensional
behavior also seems to appear near the singularity of two dimensional
black hole \foot{A complexified framework could be useful for this
$SL(2,\IR)/U(1)$ coset model as well.} \blkd.

  The topological $G/G$ theories can also be placed in the space of
topological two dimensional Landau$\,  $-\-$\,  $Ginzburg (LG)
theories \lgf. They appear at the special points of the parameter
space of twisted $N=2$ models, where the structure constants are
totally symmetric as it is expected for the fusion rules $N_{ijl}$.
The fusion ring is a deformation of the LG chiral ring \ref\glg{D.
Gepner, {\it Fusion Rings and Geometry}, Santa Barbara preprint
NSF-ITP-90-184 (1990).}. Another curiosity pointed out in \lgf\ is
that different $G/G$ theories reside in the same topological LG (TLG)
space, like $SU(2)_k/SU(2)_k$ and $U(1)_{k+1}/U(1)_{k+1}$. We would
like to use this LG singularity approach to access non$\, $-\-$\,
$diagonal ADE invariants in $G/G$.

  The $G/G$ points may turn out to be helpful for the understanding of
Topological LG theories and their TLG space. A particularly important
issue is the one of ``untwisting'' \ie\  relating the topological
theories to ordinary $N=2$ supersymmetric theories. This challenge is
easily met by the twisted $N=2$ super-\-conformal (ADE) theories at
``the origin'' of the TLG space\ref\egi{T. Eguchi and S.K. Yang
{Mod.Phys.Lett.} {\bf A5} (1990) 1693. }$\, $\dvv. At this point the
``untwisted" super-\-conformal theory, is given as a conformal ``fixed
point". Its vicinity in the TLG space, seems to be the twisting of the
``flow" into the $N=2$ superconformal theory. The theories close to
``the origin", seem therefore, to untwist into the non-\-conformal
theories obtained by the deformations of the $N=2$  superconformal
theories. We can now turn to the $G/G$ points, to look for more hints
concerning the nature of these non-\-conformal theories.  $G_k/G_k$ is
a conformal topological field theory. In its CBRST formulation, $G/G$
looks like the twisting of the $N=2$ super-\-conformal theory of
$G^c$.  It would be instructive to find directly the $N=2$
supersymmetric theory, involving the $G^c_k$ degrees of freedom, which
is twisted into $G_k/G_k$. (The complexified $G^c$ is the right arena
for such an $N=2$ theory rather than the compact group $G$ although
some $N=2$ cosets are fine \ref\kzk{Y. Kazama and H. Suzuki,
Nucl.Phys. {\bf B321} (1989) 232 and Phys.Lett. {\bf 216B} (1989)
112.}$\,$). The detailed construction is, however, still missing. The
only way we know, so far, to untwist $G_k/G_k$  is by first bosonizing
it \denm. It would be interesting to probe the TLG space around its
$G/G$ points and to relate it to the ``flow" into the  fixed point of
the $N=2$ $G^c_k$ theory. The whole subject of the relationship
between ``flows" in the topological space (as described in the LG
approach) and the flows associated with perturbations of the $N=2$
superconformal theories is an important open problem. In particular,
it would be important to determine those deformations which are
integrable. Some work in this direction was carried out recently by
Nemeschansky and Warner \ref\new{D. Nemeschansky and N.P. Warner, {\it
Topological Matter, Integrable Models and Fusion Rings} USC preprint
USC-91-031 (1991).}. The $G/G$ theories may provide a good starting
point for finding the integrable deformations. The observation of
Witten that deformations of CS theories in three dimensions lead to
integrable models in two dimensions  \ref\wii{E. Witten, Nucl.Phys.
{\bf B322} (1989) 629 and Nucl.Phys. {\bf B330} (1989) 285.}, combined
with the close relation of $G/G$ to CS theories, may provide an
important clue for locating the integrable deformations.

  It is interesting to note that $G_k/G_k$ amplitudes can be obtained
also form a discretized approach starting with a triangulation and
going over to the dual $\phi^3$ graph\foot{This observation was made
in discussions with A.A. Migdal and Y. Sonnenschein.}. We associate
with each vertex of this graph the numbers $\{N_{ijl}\}$  satisfying
duality $N_{ijk}\, N_{klm} = N_{iln}\, N_{njm}$ and completeness
$N_{ilk} N_{klj}= \delta_{ij}$. For a fixed area (and genus) all the
triangulations are obtained by the ``flip" operation \ref\mig{A.A.
Migdal, Int.Jour.Mod.Phys.  {\bf C1} (1990) 165. }. Moreover, since
duality and completeness allow to take away palquettes from the graph,
all configurations with the same genus are equivalent. We are left,
therefore, with one simple graph per genus. In particular there is no
dependence on the area, as indeed should be expected for a topological
theory. Specifying the boundary of the triangulated graph or
equivalently the set of incoming legs $\{i_1, \,i_2, \,\ldots,
\,i_{\nh}\}$ in the dual $\Phi^3$ graph, leads (on the sphere) to the
${\nh}$ point function \eqn\npf{A_{\nh}=\sum_{ \{k_i\} } N_{i_1i_2k_1}
\,N_{k_1i_3k_2} \,\ldots\, N_{k_{{\nh}-3}i_{{\nh}-1}i_{\nh}},} which
is precisely the result of $G_k/ G_k$\giii. The fusion rules of $G_k$
provide an example for  ${N_{ijl}}$'s which satisfy the duality and
completeness requirements mentioned above and are symmetric as well.

  Actually, the ${N_{ijl}}$'s associated with any CFT could do just
the same job. We focus our attention to the  RCFT case since in this
case the states associated with the links belong to a finite set (the
range of the index $i$). In particular we look at $G/H$ as a generic
example. We would thus expect to be able to define a topological
theory based on the coset model $G/H$ which would naturally be called
$(G\!/\!H)\,/ \,(G\!/\!H)$. The formalism for this kind of iterated
cosetting was presented in \witc. \me It has the form of an
inclusion-exclusion procedure, which is cohomological in its nature.
We are now looking for a framework, based on CBRST cohomology, for
$(G\!/\!H)\,/\,(G\!/\!H)$ theories. We also try to  formulate them as
gauge theories. We hope to be able to show that the amplitudes are
given in terms of the  ${N_{ijl}}$'s following from the $G/H$ model,
as expected. A better understanding of the amplitudes directly from
the CBRST cohomology will be most useful here, as well.

  Next, we would like to discuss some applications of $G/G$ theories.
Our investigation of topological cosets is helpful for general $G/H$
coset models (by far the most common construction for RCFTs). The
mechanism of the $G/H$ cosetting can be presented along the lines of a
$H/H$ theory for $H_{k^{'}} \subset G_k$ (and $k^{'}$ properly taken).
The $H_{k^{'}}$ degrees of freedom are provided by $H \subset G$
\gak$\, $\karb$\, $\kara. Adding to the $G$ theory the extra degrees
of freedom$\, $--$\, $the $H^c/H$ current algebra and the complex
ghost system in the adjoint representation of $H$, turns the $H
\subset G$ part of the $G$ model into a standard $H/H$. The CBRST
cohomology (with respect to the $H$ current algebra), leads to $G/H$.

  It should be also noted that $H/H$ theories present a potential
ambiguity in the context of $G/H$ coset models (and in principle, may
find use in the understanding of $G_k$ itself and the Quantum Group
structure found in its amplitudes \ref\kz{V. Knizhnik and A.B.
Zamolodchikov, Nucl.Phys. {\bf B247} (1984) 83.}\zam). A demonstration
for this ambiguity can be found in the context of the stringy two
dimensional black hole \blkd\ref\mbl{M. Spiegelglas,  {\it String
Winding in a Black Hole Geometry},  Technion preprint, PH-23-91.}. An
extra set of winding around a black hole was found for the string,
following from a $U(1)/U(1)$ ambiguity in the $SL(2,\IR) /U(1)$ (or
the $SU(2)/U(1)$ for region III) coset model. In \mbl, the analogy to
the statistical mechanics $Z_k$ clock model was used to sort this
ambiguity.

  We have already mentioned in section 3 $SU(2)_1/\IR$ as the way to
get the discrete states of $c_m=1$\foot{ $c_m$ denotes here, the
Virasoro anomaly of the matter system coupled to two dimensional
gravity. This system may be described by a topological matter system
with $c=0$ coupled to topological gravity. This $c=0$ system may be in
its turn, a twisted version of an $N=2$ superconformal theory with
$c_u>0$. In the example of the twisted minimal $N=2$ models
\ref\kkl{K. Li, Nucl.Phys. {\bf B354} (1991) 711 and 725.} $c_u={3\,k
\over k+2}$  These models correspond to the one-matrix models, which
in turn describe two dimensional gravity coupled to the $(2,2 k+1)$
minimal model matter system with $c_m=1-{3 (2k- 1)^2 \over  2k+1}$}.
We have also discussed the analogy of $G/G$ to this theory, both in
the formal structure (discussed in section 1) and in the appearance of
``discrete" states in $\ng\not=0$ sectors of the cohomology (in
section 4). We would also like to address briefly the issue of
coupling the topological $G/G$ theories themselves to gravity. We have
started this project motivated by the hope to find topological $c=0$
matter systems which when coupled to topological gravity, give the
$c_m<1$ (in particular, unitary) matter theories coupled to gravity.  This
question is still under investigation.  Other examples of this type
were given in ref. \kkl, where the topological theories given by
twisted minimal $N=2$ models \egi$\, $\dvv\ were coupled to topological
gravity and found to describe some minimal models coupled to gravity. It
should be noted that the models of \kkl\ are actually in the same LG
space as $SU(2)_k/SU(2)_k$ and $U(1)_{k+1}/U(1)_{k+1}$ \lgf. We
would like to couple $G/G$ to topological gravity and understand it as
some $c_m$ matter system coupled to two dimensional gravity. This may
be very tricky, since enumerating $G/G$ (and $(G\!/\!H) \,/ \,(G\!/
\!H)$ as well) theories, seems to show more theories than needed to
represent $c_m \leq 1 $ matter coupled to gravity if one would expect
to find minimal models (of course there many more models, since we
could always tensor whatever $c_m<1$ model we have with our favorite
$c=0$ model without changing its $c_m$). Should this be the case, one
is led to suspect that either some of the theories represent $c_m>1$
matter when they couple to gravity (and hence, their simplicity and
tight algebraic structure, could bring some insight to the difficult
$c_m>1$ domain of two dimensional gravity) or that some different
topological theories become identical when coupled to gravity (the LG
spaces may hint in this direction). Hopefully understanding this issue
of $G/G$ theories coupled to topological gravity as matter coupled to
gravity, will bring some insight into the question--what does gravity
couples to, after all.

  We would like to abstract from the structure of the $\nh$ points
functions in $G/G$ theories, an algebraic concept which is related to
the topology of the world-sheet. The formal structure of the
cohomology, is presented in terms of ``bounadries" \bndr\ of
multiplets. The amplitudes on the other hand, are given by the fusion
rules which are calculated using ``boundaries'' and their ``filling''.
It takes $\nh-3$ filling to add $\nh$ spins or to get the fusion rules
relevant to the $\nh$ points function (as seen in \npf). ``Filling"
is also employed for higher genus amplitudes (applied $2(g-1)$ time)
and constitutes, therefore, the (non-abelian) group theoretical
concept corresponding to the topology of the world-sheet. It remains
to be seen whether ``filling" could be a  clue for the coupling the
$G/G$ theory to geometry and gravity.

  \bigskip

{\bf~~Acknowledgments:}  It is a pleasure to thank D.
Bar-Natan, M. Bershadsky, S. Elitzur, E. Frenkel, K. Gaw{\c{e}}dzki,
V. Kac, M.S. Marinov, A.A. Migdal, D. Nemeschansky,  A.M. Polyakov, N.
Sochen, Y. Sonnenschein, C. Vafa, E. Witten and A.B. Zamolodchikov,
who shared with us their insights and criticism during the last two
years, while these ideas were shaped and written. The research was
partially supported by the Technion V.P.R. Fund, the C. Wellner
Research Fund, the US Israel Binational Science Foundation (BSF)  and
the Israeli Academy of Sciences and Humanities. M.S. is also grateful
to the Institute for Advanced Study, Princeton for a visit during the
completion of this work.

\Appendix{} {Complex Gauge Transformations in Real Notation.}

  We want to examine the significance of complex gauge transformations
and provide an introduction to the subject \atbt. We start with a real
abelian gauge field $\vec {A}= (A_x,A_y)$. We previously used $\az =
A_x \, + \, iA_y$ and $\azb = A_x \, - \, iA_y$. In a topologically
trivial situation (taking here the world-sheet to be $\WS=\RN^2$) let
us choose the gauge $\vec{\nabla}\cdot \vec{A}=0$. $\vec{A}$ could now
be written as $(\pr{y} \im,\,-\pr{x} \im)$. For a given field strength
$F$, $\im$ satisfies $\nabla^2 \im=F$ (when $\WS$ is compact, the
obstruction for the solution of this Poisson equation is the first
Chern class of the connection $\vec{A}$.). If we do not wish to choose
a particular gauge we could take $\vec{A}=(\pr{x} \re \, +\,\pr{y}
\im,\; \pr{y}\re \,-\, \pr{x} \im)$. In complex notation we take
$\lambda\,=\, \re\, + \, i \im$, $\bar{\lambda}\,=\,\re\, - \, i \im$
and have $ \az= \pz \lambda$ and $\azb= \pzb \bar{\lambda}$. This
allows to write the gauge configuration $\vec{A}$ as a complex gauge
transformation of $\vec{A}=\vec{0}$. We use $\im$, the imaginary part
of the gauge transformation, to encode $F$. In the nonabelian case
$h(z)$ or more precisely $hh^*$, plays the same r\^ole. In that case
$hh^*$ is analogous to $\im = {\rm im } \lambda$. The ``rest" of
$h(z)$ (encoded less canonically in $f=\sqrt{hh^{*}} h^{-1}\,$) is a
real gauge transform on $g$ and $\az$ much like $\re={\rm re }
\lambda$.

  We were careful to use the term gauge transformations rather than
gauge symmetries since only $\re={\rm re } \lambda$ generates a
symmetry of the action. $\im={\rm im} \lambda$ becomes a dynamical
variable, enabling us to trade the gauge field physical degrees of
freedom for the scalar field $\im$, denoted by $Y$ in section 2 and
given by $hh*$ in the WZW case. The matter fields ``feel" only the
real gauge transformations. This sets the stage for nontrivial $\WS$'s
with topologically non-\-trivial gauge configurations, resulting from
holonomies around handles and holes. All this global information is
encoded in $\re$, the real part of the gauge transformations. The
allowed $\re$ depends on the matter system coupled to $\vec{A}$ and
should be taken to be multivalued in order to change holonomies. The
topological information is insensitive to imaginary gauge
transformations, which merely deals with the local curvature. This
separation between local and global information, makes complex gauge
transformations so useful in setting the stage for the study of the
topology hiding in gauge configurations. A topologically equivalent
flat gauge configuration can be reached from any gauge configuration
via a complex gauge transformation.

\listrefs
\bye